\documentclass[a4paper,11pt]{article}
\usepackage{amsfonts}

\usepackage{graphicx}
\usepackage{amsmath}
\usepackage{amssymb}
\usepackage{latexsym}
\usepackage[colorlinks]{hyperref}
\usepackage{color}
\usepackage{float}
\usepackage{cite}
\usepackage{makeidx}
\usepackage{colortbl}
\usepackage{csquotes}
\usepackage[squaren]{SIunits}

\usepackage[textfont={footnotesize,sf},labelfont={color=blue,bf,sf},labelsep=endash]{caption}
\usepackage[position=top,labelfont={color=blue,bf,sf}]{subfig}
\usepackage{bm}

\usepackage[title]{appendix}

\newcommand{\bra}{\begin{array}}
\newcommand{\era}{\end{array}}
\newcommand{\beq}{\begin{equation}}
\newcommand{\eeq}{\end{equation}}
\newcommand{\beqar}{\begin{eqnarray}}
\newcommand{\eeqar}{\end{eqnarray}}

\def\BC{\bb C}
\def\_\BC{\bbi C}

\def\( {\left(}
   \def\) {\right)}
\def\[ {\left[}
\def\] {\right]}
\def\no2 {{\textstyle{n\over 2}}}

\begin{document}
\begin{titlepage}
\setcounter{page}{1}
\renewcommand{\thefootnote}{\fnsymbol{footnote}}
\begin{flushright}
\end{flushright}
\vspace{5mm}
\begin{center}
{\Large \bf {Energy  Levels of Gapped Graphene Quantum Dot in
Magnetic Field}}

\vspace{5mm}

{\bf Abderrahim Farsi}$^{a}$, {\bf Abdelhadi Belouad}$^{a}$ and  {\bf
Ahmed Jellal\footnote{\sf 
a.jellal@ucd.ac.ma}}$^{a,b}$

\vspace{5mm}

{$^{a}$\em Laboratory of Theoretical Physics,  
Faculty of Sciences, Choua\"ib Doukkali University},\\
{\em PO Box 20, 24000 El Jadida, Morocco}

{$^{b}$\em Canadian Quantum  Research Center,
	204-3002 32 Ave Vernon, \\ BC V1T 2L7,  Canada}

\vspace{3cm}

\begin{abstract}
We study the energy levels  of carriers confined in a magnetic quantum dot of graphene surrounded by a infinite graphene sheet in the presence of energy gap.
The eigenspinors are derived for the valleys $K$ and $K'$, while the associated energy levels are obtained by using the boundary condition at interface of the quantum dot. We numerically investigate our results and show that the energy levels exhibit the  symmetric and antisymmetric behaviors under suitable conditions of the physical parameters. We find that the radial probability can be symmetric or antisymmeric according to the angular momentum is null or no-null. Finally, we show that the application of an energy gap decreases  the electron density in the quantum dot, which indicates a temporary trapping of electrons.

\vspace{5cm}

\end{abstract}
\end{center}
\end{titlepage}

\section{Introduction}

Graphene is a two-dimensional crystalline material that is an allotropic form of carbon
and the stack of which constitutes graphite \cite{K. S. Novoselov04}.
Due to its special properties, graphene has recently been attracted by considerable attention \cite{A. H. Castro Neto09 , Y. Zhang05 , G. Jo10}. It was prepared using several techniques, including surface precipitation of silicon carbide \cite{C. Berger06 , T. Ohta06}. 
  In the context of band theory, graphene appears as a special case since the valence and conduction bands are touched at two Dirac points   $K$ and $K'$ (valleys) defining  the edge of first Brillouin zone.
  It is 
   characterized by   a linear  dispersion relation  in contrary to  semiconductors, which have  parabolic ones. This behavior makes possible to look at 
   electrons in graphene as relativistic particles with a zero effective mass and 
   having a velocity of the effective light called Fermi velocity   of order of $10^6$ m/s.

 In recent years, quantum dots (QDs) in graphene have been the subject of intensive research due to their unique electronic and optical properties \cite{Rozhkov11,Trauzettel07}. The QD of single layer graphene  contains small cut flakes in which the confinement of the support is due to the quantum size effect. Electrostatic confinement of electrons in integrable graphene
QDs have also been proposed in which the edge effect is no longer significant \cite{Bardarson09}. Their electronic and optical properties  depend on the shape and edges.
 For example, in the presence of zigzag edges, the energy spectrum of  QD has zero energy levels, while with a wheelchair the spectrum shows an energy deficit \cite{Zhang08,Zarenia11,Ezawa07}.
 In the absence of a spectral gap it was theoretically shown that an electrostatically confined QD can accommodate only quasibound states  \cite{Matulis08,Hewageegana08}.
 Recent theoretical and experimental results have shown that a gap can be induced in graphene by modifying the density of charge carriers via the application of an external field or  chemical doping, which creates a potential difference 
 \cite{Ohta06,Min07}.
 The energy levels of circular graphene QDs in the presence of a perpendicular magnetic field were recently investigated analytically for the special case of infinite mass boundary condition \cite{Schnez08}.
 A periodic magnetic field applied perpendicular to the graphene  can preserve the isotropic Dirac cones of the  energy bands while reducing the slope of the Dirac cones \cite{Snyman09}.

We study the charge carriers confinement in a circular QD in graphene surrounded by a graphene sheet with an external magnetic field in the presence of an energy gap  $\Delta$.
We solve  the Dirac equation to obtain the eignspinors inside and outside the QD of graphene.
By applying the boundary condition at  interface, we obtain an equation describing the energy levels in terms of physical parameters characterizing our system. We numerically study the energy levels  as a function of the angular momentum, radius of  QD, magnetic field and energy gap.
We find that the energy levels show different behaviors, which can be symmetric or antisymmetric depending on the sign of  parameters. 
We analyze a  limiting case by giving explicitly the expression of energy levels
in terms of two quantum numbers. Subsequently, we investigate the radial probability and obtain  two symmetries corresponding to angular momentum $m=0$
and $m\neq 0$. 
In addition,
we study  the electron density and show that its increase  in the quantum dot indicates a temporary trapping of electrons. We conclude that the energy gap can be used as a tunable parameter to control the electronic properties of our system.

The paper is organized as follows. In section $2$, we set our theoretical model
and determine 
the eigenspinors. Using   boundary condition, we derive a formula governing the the energy levels as function
of the physical parameters.
We numerically analyze the energy levels, radial probability and  electron density  under various conditions in section $3$. We conclude our results in the final section.

\section{Theoretical model}

\begin{figure}[h!]
  \center
  \includegraphics[width=7cm,height=4cm]{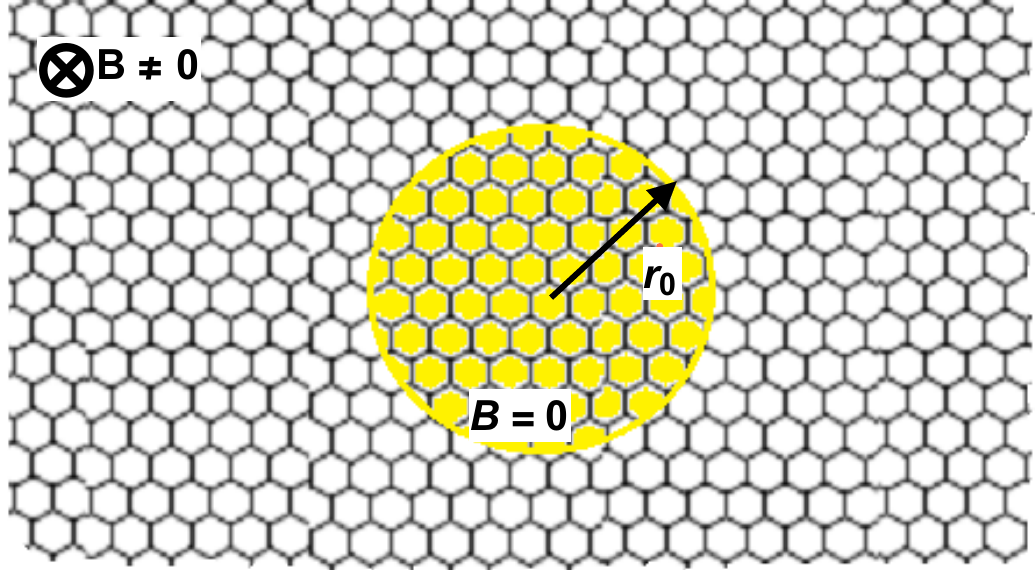}
  \caption{\sf (color online) Schematic diagram of a circular quantum dot  of radius $r_{0}$  surrounded by a graphene
sheet in the presence of a perpendicular magnetic field $B$ outside the quantum dot.
  \label{fig1}}
\end{figure}

We consider a quantum dot (QD) of radius $r_{0}$ in graphene surrounded by an infinite graphene sheet with a non-zero magnetic field outside and zero inside QD as shown in  Figure \ref{fig1}. More precisely, 
our system can be  modeled as a circularly symmetric QD by
using an external magnetic field along $z$-direction defined by
\begin{equation}\label{e3}
\vec{B}=
\left\{
\begin{array}{ll}
B, & \hbox{$> r_{0}$} \\
0, & \hbox{$r< r_{0}$}
\end{array}
\right.
\end{equation}
which gives rise to the vector potential
\begin{equation}\label{e4}
\vec{A}=
\left\{
\begin{array}{ll}
\frac{B}{2r}(r^2-r_{0}^2), & \hbox{$r> r_0$} \\
0, & \hbox{$r< r_0$}.
\end{array}
\right.
\end{equation}
To describe the dynamics of carriers in the
honeycomb lattice of covalent-bond carbon atoms of gapped graphene, 
we introduce  the 
Hamiltonian
\begin{equation}\label{e1}
H_\eta=v_F\left(\pi_x\sigma_x+\eta\pi_y\sigma_y\right)+\Delta\sigma_z
\end{equation}
where $v_F = 10^6$ m/s is the Fermi velocity, $\pi_i=p_i+eA_i$ are the conjugate momentum, $\sigma_i$ are the Pauli matrices in the
basis of the two sublattices of $A$ and $B$ atoms, $\eta=\pm1$ labels the  valleys  $K$ and $K'$,  $\Delta $ is the energy gap. 
In 
the polar coordinates $(r,\varphi)$,
the Hamiltonian \eqref{e1} takes the form
\begin{equation}\label{e5}
H_\eta=\hbar v_F
\begin{pmatrix}
\Delta &  e^{i\eta\varphi}\left[-i \frac{\partial}{\partial r}+i\eta \left(-\frac{i}{r} \frac{\partial}{\partial\varphi}+ \frac{eA\varphi}{\hbar}\right)\right] \\
e^{-i\eta\varphi}\left[-i\frac{\partial}{\partial r}-i \eta \left(-\frac{i}{r} \frac{\partial}{\partial\varphi}+\frac{eA\varphi}{\hbar}\right)\right] & -\Delta \\
\end{pmatrix}
\end{equation}

 Due to the circular symmetry, 
 the Hamiltonian $H_\eta$ commutes with 
 the total angular momentum $J_z=-i\hbar\partial_\varphi+\hbar\sigma_z/2$. This implies that
  the eigenspinors can be separated as
\begin{equation}\label{e2}
\Psi(r,\varphi)=
                      \begin{pmatrix}
                        \psi_{A,\eta}(r,\varphi) \\
                        \psi_{B,\eta}(r,\varphi) \\
                      \end{pmatrix}
                   =
e^{im\varphi}
\begin{pmatrix}
  R_{A,\eta}(r) \\
  e^{i\eta\varphi}R_{B,\eta}(r) \\
\end{pmatrix}
\end{equation}
where $m=0,\pm1,\pm2 \cdots $ are eigenvalues of  $J_z$.
To determine  the radial components $ R_{A,\eta}(r)$  and $R_{B,\eta}(r)$,
we use   the eigenvalue equation $H_\eta\Psi=E\Psi$ to get
\begin{eqnarray}
&&\label{e8}
  \hbar v_F e^{-i\eta\varphi}\left[-i\frac{\partial}{\partial r}-i\eta\left(\frac{m}{r}+\frac{eB}{\hbar r}\left(r^2-r_0^2\right)\right)\right]R_{B,\eta}(r)=\left(E-\Delta\right)R_{A,\eta}(r)\\
  &&
  \label{e88}
  \hbar v_F e^{i\eta\varphi}\left[-i\frac{\partial}{\partial r}+i\eta\left(\frac{m}{r}+\frac{eB}{\hbar r}\left(r^2-r_0^2\right)\right)\right]R_{A,\eta}(r)=\left(E+\Delta\right)R_{B,\eta}(r)
\end{eqnarray}
 By injecting  \eqref{e88} into \eqref{e8}, we obtain a
differential equation for $R_{A,\eta}$
\begin{equation}\label{e888}
\left[-\frac{\partial^2}{\partial r^2}-\frac{1}{r}\frac{\partial}{\partial r}+ \frac{m_{\text{eff}}^2}{r^2}+\frac{1}{4}\left(\frac{eB}{\hbar}\right)^2 r^2+\frac{eB}{\hbar}\left(m_{\text{eff}}+\eta\right)-\frac{E^2-\Delta^2}{\left(\hbar v_F\right)^2}\right]R_{A,\eta}=0
\end{equation}
where we have set the quantum number  $m_{\text{eff}}=m-s$ such that  $s={B \pi r^2_{0} e}/{h}$ is the "missing" flux and indicates the amount of magnetic flux screened out from the magnetic QD \cite{Myoung19}.

In the forthcoming analysis, we introduce the dimensionless units
$E_0=\frac{\sqrt{2}\hbar v_F}{l_B}$,   $\varepsilon=\frac{E}{E_0}$,  $\delta=\frac{\Delta}{E_0}$ and the variable change $\rho=\frac{r}{l_B}$, with  $l_B=\sqrt{\frac{\hbar}{e B}}$ is the magnetic length.
  Now, to get the solution of energy spectrum, we solve \eqref{e888} for each region composing our system. For region $r< r_0$, \eqref{e888} reduces to
\begin{equation}\label{e9}
\left[\frac{\partial^2}{\partial \rho^2}+\frac{1}{\rho}\frac{\partial}{\partial \rho}-\frac{m^2}{\rho^2}+\alpha^2\right]R_{A,\eta}=0
 \end{equation}
which has  Bessel function of the
first kind as solution
and therefore the first component of eigenspinor takes the form
\begin{equation}\label{e100}
    \psi_{A,\eta}(\rho,\varphi)= C_{1}e^{im\varphi} J_{|m|}\left(\alpha\rho \right)
\end{equation}
where $C_1$ is the normalization constant and we have defined $\alpha= \sqrt{2|\varepsilon^2-\delta^2|}$. The second component of eigenspinor can be obtained from \eqref{e88} by using \eqref{e100} to end up with
\begin{eqnarray}\label{e11}
    \psi_{B,\eta}(\rho,\varphi) =-i \frac{C_1}{2}e^{i(m+\eta)\varphi}
    \left[
    \sqrt{\left|\frac{\varepsilon-\delta}{\varepsilon+\delta}\right|}
    \left(J_{|m|-1}(\alpha\rho)-J_{|m|+1}(\alpha\rho)\right)
    -\frac{\eta m\sqrt{2}}{\rho}J_{|m|}(\alpha\rho)
    \right].
\end{eqnarray}
For region $r> r_0$, we write \eqref{e888} in dimensionless units as
 \beq \label{e17}
 \left( \frac{\partial^2}{\partial
\rho^2}+\frac{1}{\rho}\frac{\partial}{\partial \rho}-\frac{m_{\text{eff}}^2}{\rho^2}-\frac{1}{4}\rho^2-m_{\text{eff}}-\eta+\alpha^2\right)R_{A,\eta}=0.
 \eeq
It can be solved by introducing the following ansatz
 \beq\label{e18}
 R_{A,\eta}(\rho)=\rho^{|m_{\text{eff}}|}e^{-\frac{\rho^2
}{4}}\chi(\rho^{2}) \eeq
  yielding the confluent hypergeometric ordinary
differential equation
\beq\label{e19}
\left[x \frac{\partial^2}{\partial x^2
}+(b-x)\frac{\partial}{\partial x}-a\right]\chi(x)=0
\eeq
 where we have set $x=\frac{\rho^{2}}{2}$ and the  quantities
\beq\label{e20}
 b=1+|m_{\text{eff}}|, \qquad
2a= |m_{\text{eff}}|+m_{\text{eff}}+1+\eta-\alpha^2.
\eeq
It has the confluent hypergeometric function $U\left(a,b,\frac{\rho^{2}}{2}\right)$ as solution with $C_{2}$ is the normalization constant.
Consequently, we obtain the first spinor component
 \beq \label{e21}
\psi_{A,\eta}(\rho,\varphi)=C_{2} \rho^{|m_{\text{eff}}|}e^{-\frac{\rho^{2}}{4}}U\left(a,b,\frac{\rho^{2}}{2}\right)e^{im\varphi} \eeq
and the  second component can be extracted form 
\eqref{e88} as
\begin{eqnarray} \label{e22}
\psi_{B,\eta}(\rho,\varphi) &=&
\frac{i C_{2}}{\sqrt{2}\left(\varepsilon+\delta\right)}e^{i\left(m+\eta\right)\varphi}
\rho^{|m_{\text{eff}}|-1}e^{-\frac{\rho^{2}}{4}}\\
&&
\left[\left(\eta m_{\text{eff}}-|m_{\text{eff}}|+ \frac{\eta+1}{2}\rho^2\right)U\left(a,b,\frac{\rho^2}{2}\right)+a\rho^{2}U\left(a+1,b+1,\frac{\rho^2}{2}\right)\right]\nonumber.
\end{eqnarray}
Finally,  for region $r<r_0$ the eigenspinors are
\begin{eqnarray}\label{e23}
\psi(\rho,\varphi) = {C_1} e^{im\varphi}
\begin{pmatrix}
 J_{|m|}(\alpha\rho) \\
  -i\frac{  e^{i\eta\varphi}}{\sqrt{2}(\varepsilon+\delta)} 
  \left[\frac{\alpha}{2}\left(J_{|m|-1}(\alpha\rho)-J_{|m|+1}(\alpha\rho)\right)-\frac{\eta m\sqrt{2}}{\rho}J_{|m|}(\alpha\rho)\right]
\end{pmatrix}
\end{eqnarray}
and for region $r>r_0$ we have
\begin{eqnarray}\label{ee24}
\psi(\rho,\varphi) &=& C_2 e^{im\varphi} \rho^{|m_{\text{eff}}|-1}~e^{-\frac{\rho^{2}}{4}} \\
&& \begin{pmatrix}
\rho~U\left(a,b,\frac{\rho^{2}}{2}\right)\\
\frac{i e^{i\eta\varphi}}{\sqrt{2}\left(\varepsilon+\delta\right)}\left[\left(\eta m_{\text{eff}}-|m_{\text{eff}}|+ \frac{\eta+1}{2}\rho^2\right)U\left(a,b,\frac{\rho^2}{2}\right)+a\rho^{2}U\left(a+1,b+1,\frac{\rho^2}{2}\right)\right]\nonumber
\end{pmatrix}.
\end{eqnarray}

As far as the energy levels are concerned, we apply the boundary condition at the radius $r=r_{0}$ of  quantum dot. Then, we have
\beqar
&&\psi_{A,\eta}(\rho_0,\varphi)_{\rho<\rho_0} = \psi_{A,\eta}(\rho_0,\varphi)_{\rho>\rho_0}\\
&&\psi_{B,\eta}(\rho_0,\varphi)_{\rho<\rho_0} = \psi_{B,\eta}(\rho_0,\varphi)_{\rho>\rho_0}
\eeqar
with the normalized radius $\rho{_0}=\frac{r_0}{l_B}$.
By using \eqref{e23} and  \eqref{ee24}
we explicitly obtain
\begin{eqnarray}\label{e24}
  &&C_{1} J_{|m|}\left(\alpha\rho_0\right)=C_{2} \rho_0^{|m_{\text{eff}}|}e^{-\frac{\rho_0^{2}}{4}}U\left(a,b,\frac{\rho_0^{2}}{2}\right) \\
  &&-C_1 \frac{\alpha}{2}\left(J_{|m|-1}(\alpha\rho_0)-J_{|m|+1}(\alpha\rho_0)\right)-\frac{\eta m}{\rho_0}J_{|m|}(\alpha\rho_0)  \nonumber=\\
  &&C_{2}\rho_0^{|m_{\text{eff}}|-1}e^{-\frac{\rho_0^{2}}{4}}\left[\left(\eta m_{\text{eff}}-|m_{\text{eff}}|+ \frac{\eta+1}{2}\rho_0^2\right)U\left(a,b,\frac{\rho_0^2}{2}\right)+a\rho_0^{2}U\left(a+1,b+1,\frac{\rho_0^2}{2}\right)\right].
\end{eqnarray}
It is convenient  to write the above equations  in matrix form
\begin{equation}\label{e25}
M
        \begin{pmatrix}
          C_{1} \\
          C_{2} \\
        \end{pmatrix}
     =
  \begin{pmatrix}
    m_{11} & m_{12} \\
    m_{21} & m_{22} \\
  \end{pmatrix}
         \begin{pmatrix}
           C_{1} \\
           C_{2} \\
         \end{pmatrix}
       =0
       \end{equation}
       where the matrix elements are given by
\begin{align}\label{e26}
  & m_{11}=J_{|m|}\left(\alpha\rho_0\right)
   \\
  &m_{12}=-\rho^{|m_{\text{eff}}|}~e^{\frac{-\rho^{2}}{4}}U\left(a,b,\frac{\rho_0^{2}}{2}\right)\\
  & m_{21}=-\frac{\alpha}{2}\left(J_{|m|-1}(\alpha\rho_0)-J_{|m|+1}(\alpha\rho_0)\right)-\frac{\eta m}{\rho_0}J_{|m|}(\alpha\rho_0)\\
  & m_{22}=-\rho_0^{|m_{\text{eff}}|-1}e^{-\frac{\rho_0^{2}}{4}}
  \left[\left(\eta m_{\text{eff}}-|m_{\text{eff}}|+ \frac{\eta+1}{2}\rho_0^2\right)U\left(a,b,\frac{\rho_0^2}{2}\right)+a\rho_0^{2}U\left(a+1,b+1,\frac{\rho_0^2}{2}\right)\right].
\end{align}
Consequently, the energy levels are solution of the condition
\begin{equation}\label{e27}
\det M= m_{11}m_{22}-m_{12}m_{21}=0.
\end{equation}
Since 
 \eqref{e27}
is a complicated task to analytically derive the energy levels, we use the  numerical approach to study their basic features.
To this end, we propose to analyze the behavior of the energy levels under various conditions of the physical parameters such that quantum angular momentum $m$, radius $r_0$ of the quantum dot,  magnetic field $B$ and energy gap $\Delta$.

\section{Numerical Results}\label{sec:results}

\begin{figure}[!hbt]\centering
	\center
	\includegraphics[width=6.4cm,height=4.2cm]{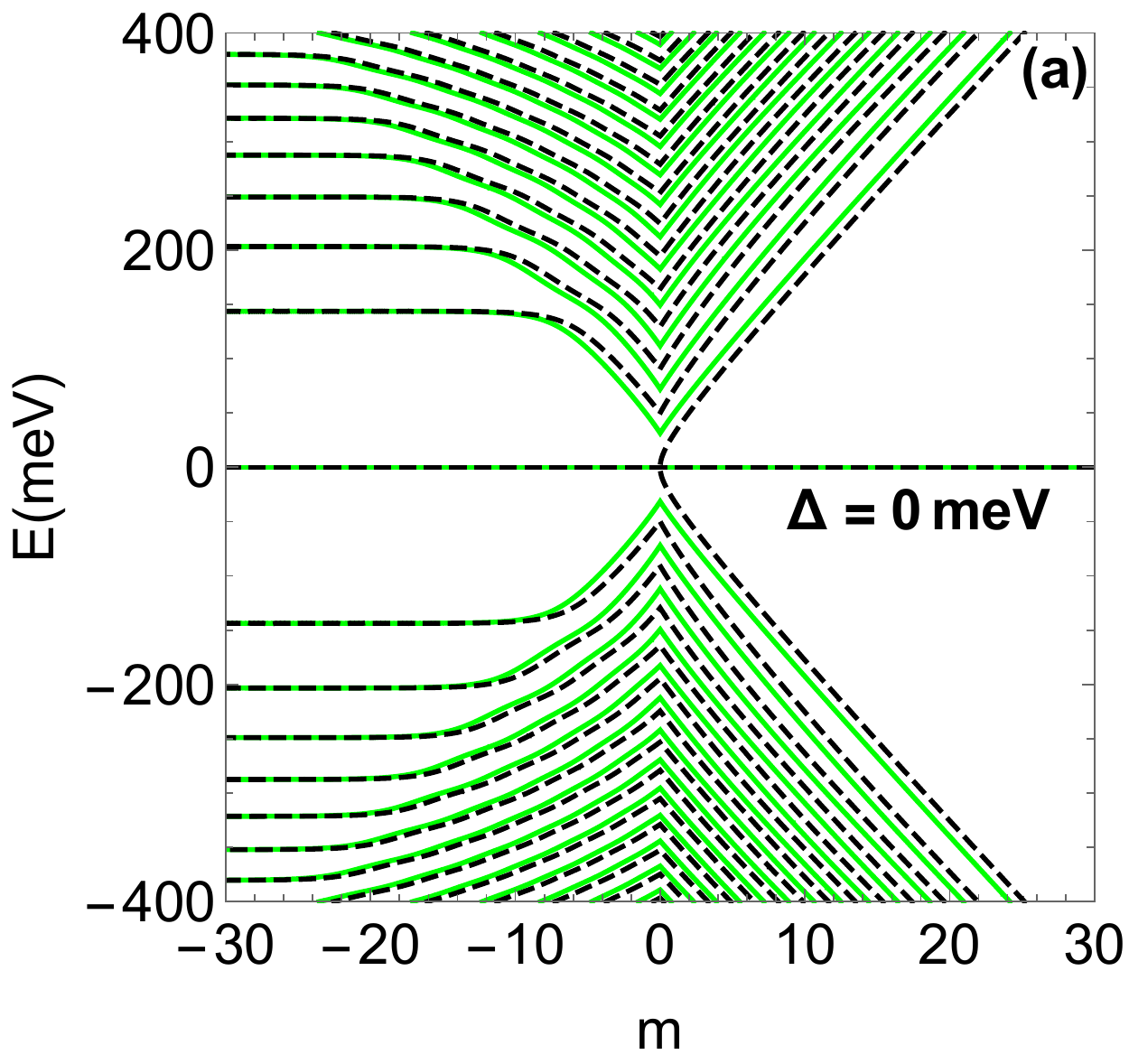}
	\hspace{2cm}
	\includegraphics[width=6.4cm,height=4.2cm]{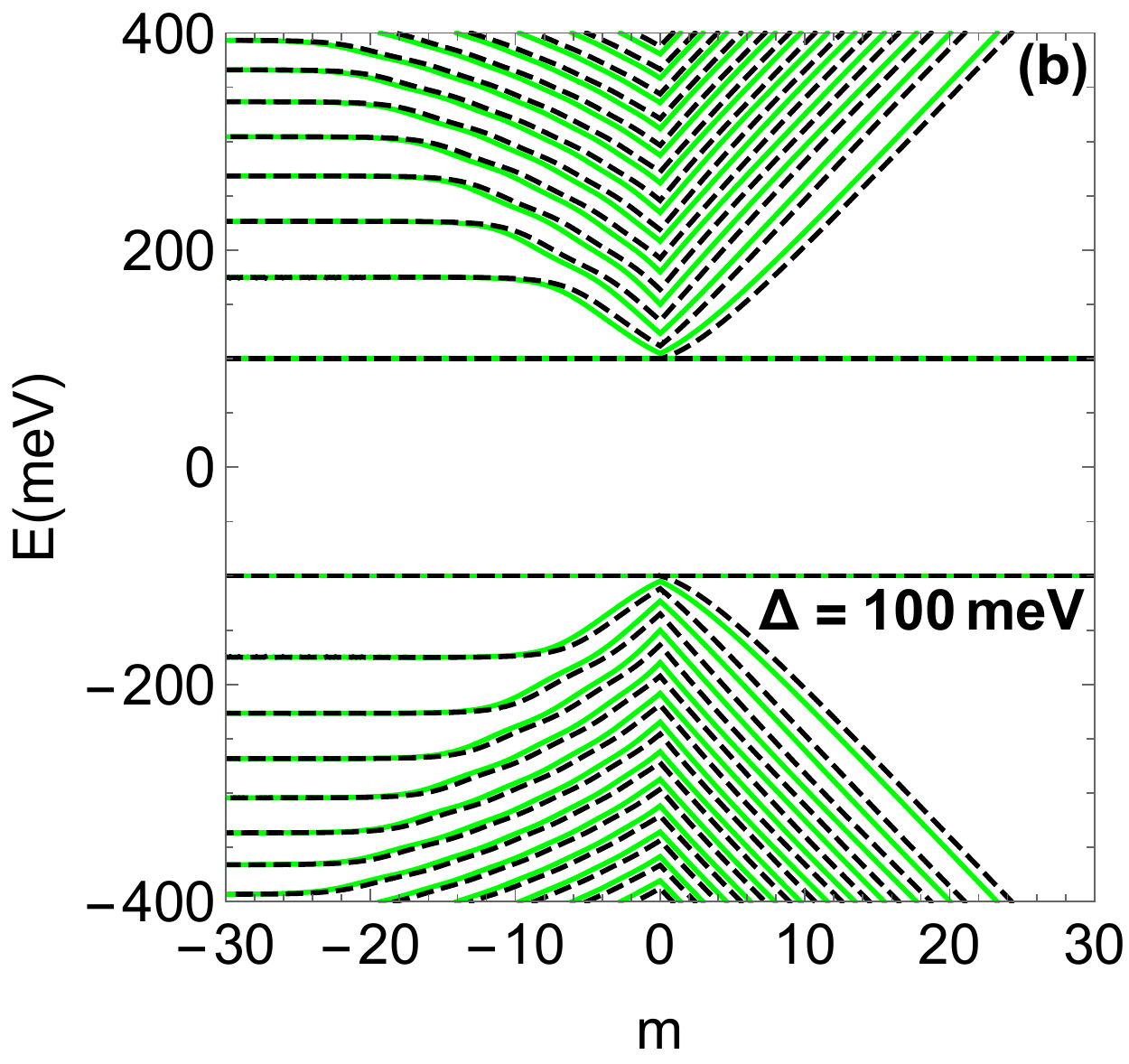}
	\caption{ Energy levels as a function of the angular momentum $m$ for  $B=15.7$ T, $r_0=44.3$ nm and two values of energy gap (a):
		$\Delta=0$ meV, (b): $\Delta=100$ meV with   green curves for $\eta=1$ and black dashed curves for $\eta=-1$.
		\label{f2}}
\end{figure}

In Figure \ref{f2}, we show the energy levels for a quantum dot in graphene as a function of the angular momentum $m$ for a magnetic field $B=15.7$ T, radius $r_{0}= 44.3$ nm and two values of energy gap such that (a): $\Delta=0$ meV  and (b): $\Delta=100$ meV. The solid  and  dashed lines correspond, respectively, to the two valleys $ K$ $(\eta=1)$ and $K'\ (\eta=-1)$. For $m\leq -10$, we observe that the energy levels  are doubly degenerate because of the symmetry  $E(m,\eta)=E (m,-\eta)$. However, this degeneracy is broken when $m> -10$, i.e. we have $E(m,\eta)\neq E(m,-\eta)$. It is clearly seen that  the energy levels present a symmetry between the valence  and  conduction bands that is $E(m,\eta)=-E (m,\eta)$.
In the case of non gap ($\Delta=0$ meV) as shown in  Figure \ref{f2}(a), we notice  the existence of zero energy
levels doubly degenerate $E(m,\eta)=E (m,-\eta)$, which are in agreement with 
those obtained in \cite{Myoung19}. Now by taking into account of energy gap ($\Delta=100$ meV), it is clear from Figure \ref{f2}(b) that 
the energy level behaviors changed and as a consequence there is  creation of a gap  ($\Delta E=200$ meV) as should be.

\begin{figure}[H]\centering
	\center
	\includegraphics[width=5.57cm,height=4.2cm]{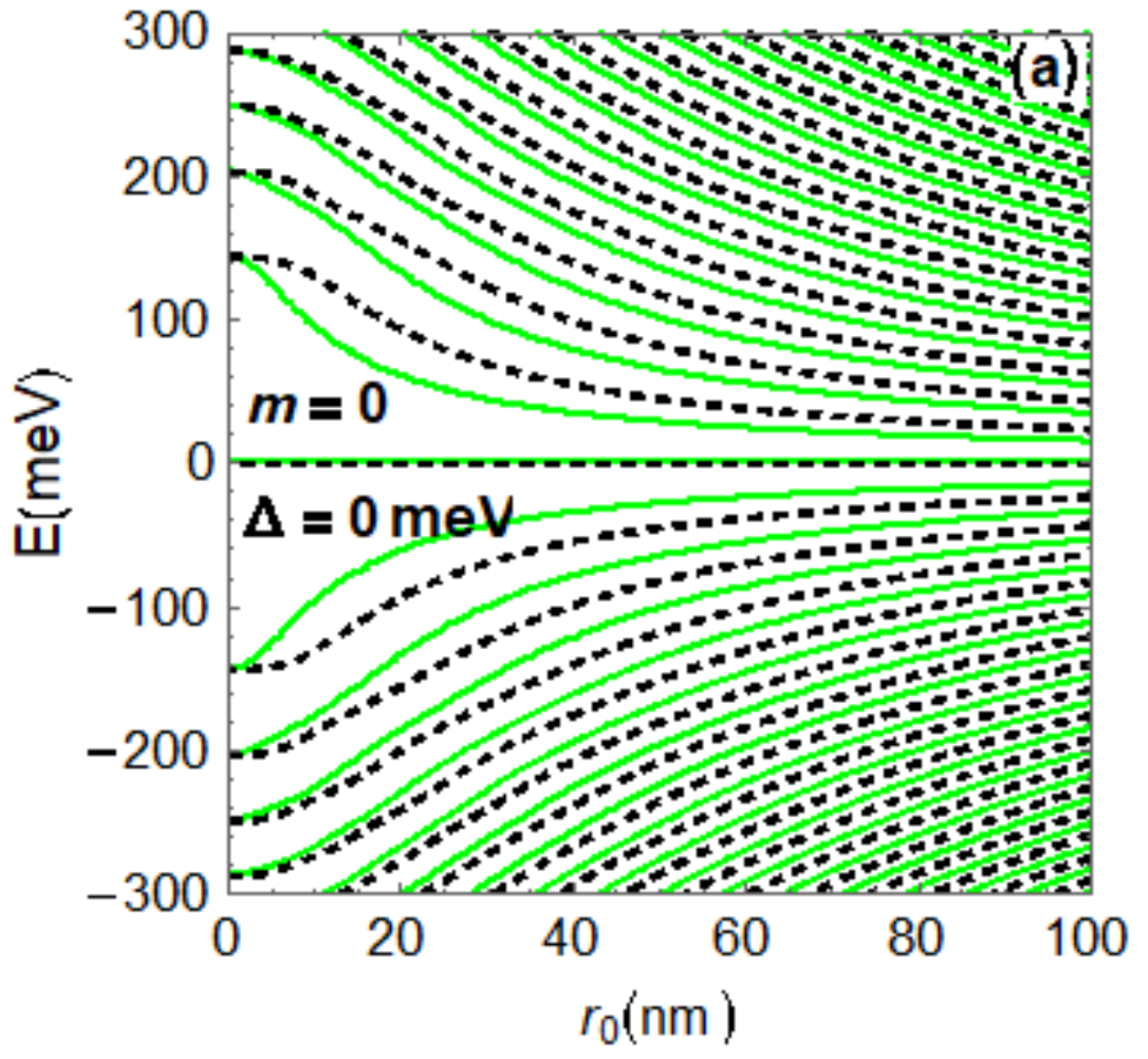}
	\includegraphics[width=5.57cm,height=4.2cm]{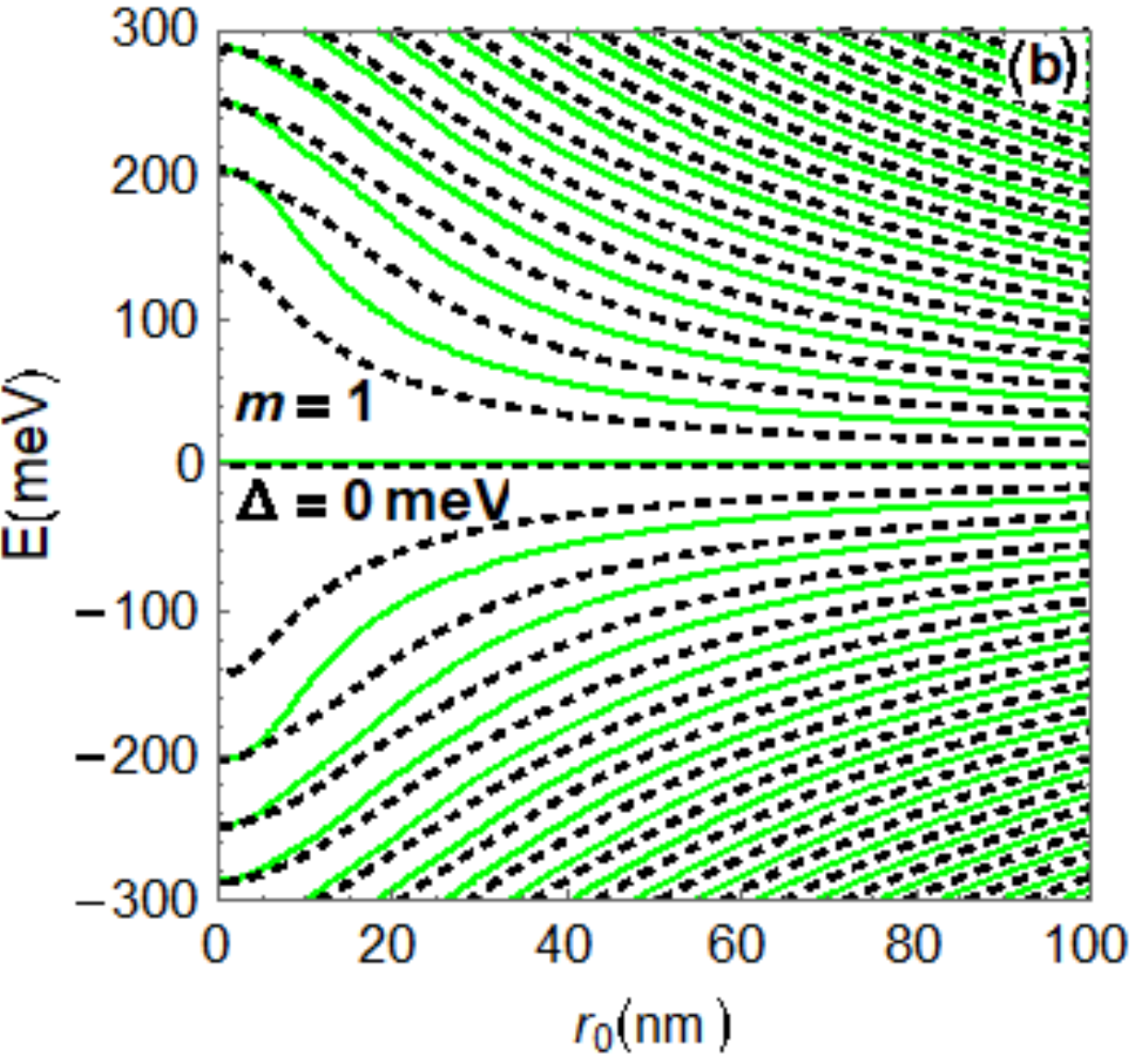} \includegraphics[width=5.57cm,height=4.2cm]{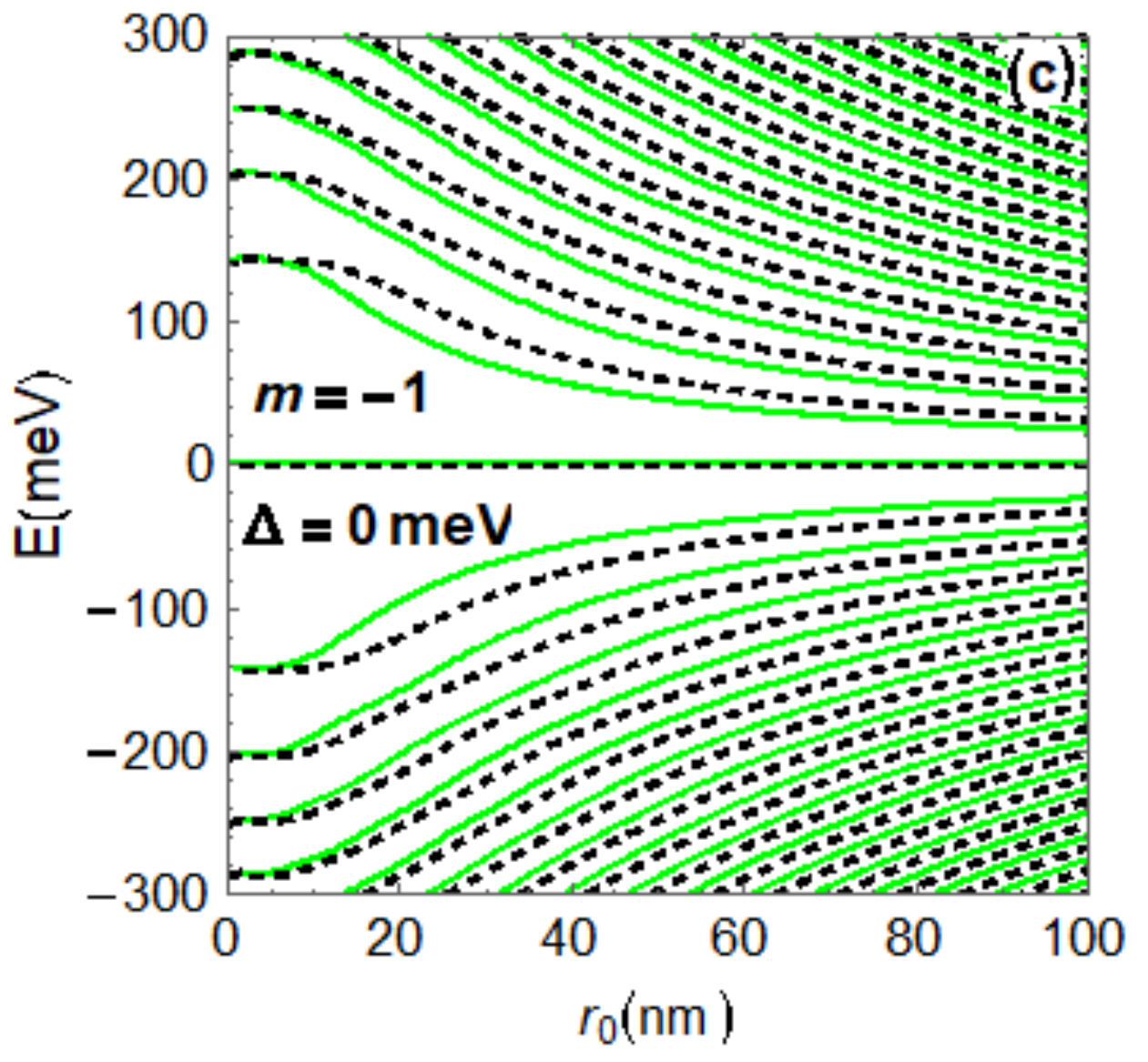}\\
	\includegraphics[width=5.57cm,height=4.2cm]{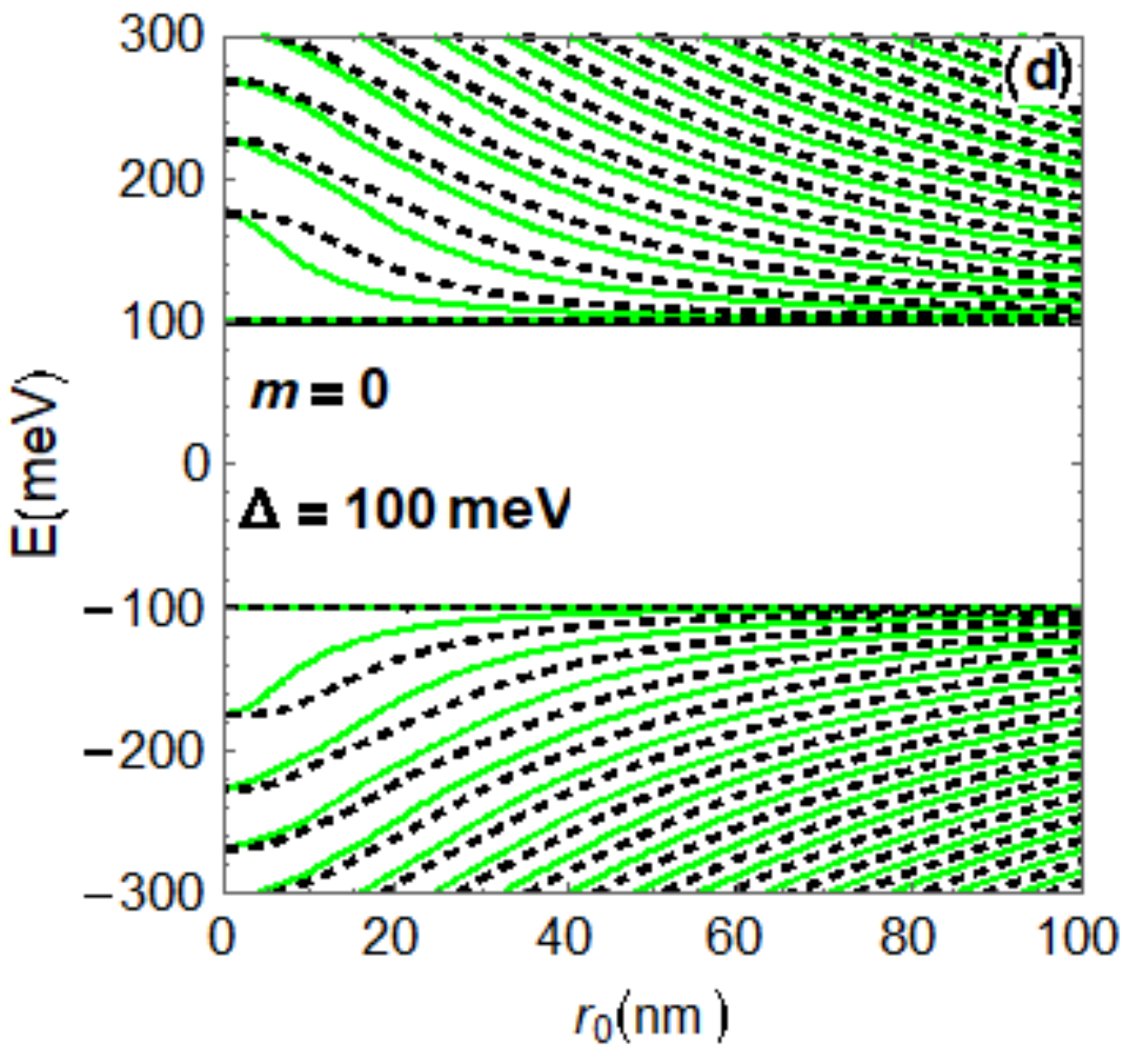}
	\includegraphics[width=5.57cm,height=4.2cm]{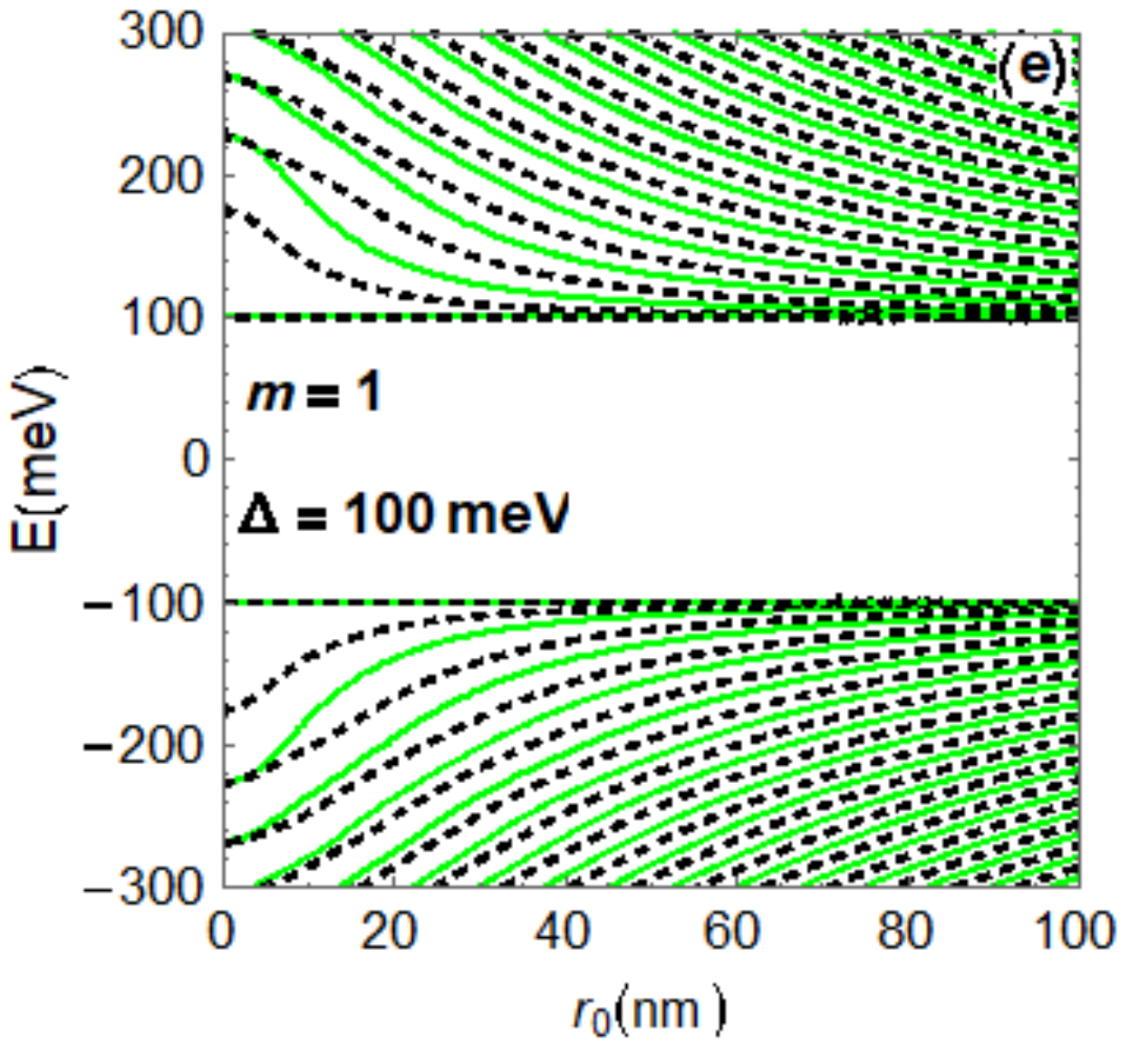}
	\includegraphics[width=5.57cm,height=4.2cm]{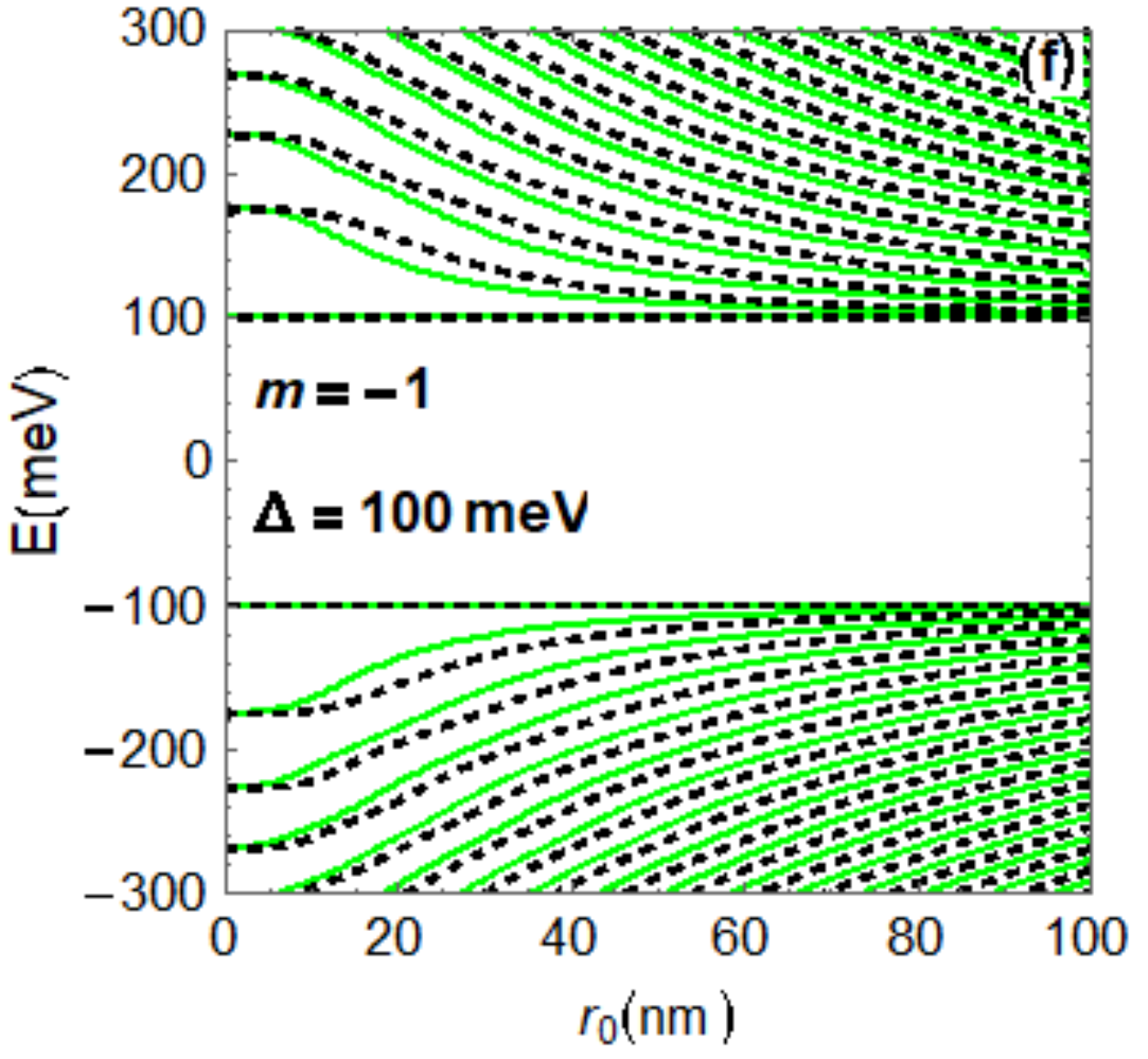}
	\caption{\sf  Energy levels as a function of the radius $r_0$ of quantum dot  for $B=15.7$ T, three values of angular momentum and two values of energy gap.  (a): $m=0$, (b): $m=1$, (c): $m=-1$ for $\Delta=0$ meV. (d): $m=0$, (e): $m=1$, (f): $m=-1$ for $\Delta=100$ meV, with   green curves for $\eta=1$ and black dashed curves for $\eta=-1$.
		\label{f3}}
\end{figure}

Figure \ref{f3} presents the energy levels as a function of the radius $r_0$ of 
quantum dot  for $B=15.7$ T, three values of angular momentum
 $m$ and two values of energy gap $\Delta$ such that  (a): $m=0$, (b): $m=1$, (c): $m=-1$
 for zero gap and  (d): $m=0$, (e): $m=1$, (f): $m=-1$ for a gap $\Delta=100$ meV. As before, the solid and dashed lines correspond, respectively to the two valleys $K \ (\eta=1) $ and $K'\ (\eta=-1)$. We observe that when $r_0$ approaches to zero, 
 the energy levels 
 are degenerated and then we have the symmetry $E (m,\eta)=E(m,-\eta)$.  When $r_0$ increases such degeneracy of the valleys $K$ and $K'$ no longer exists, which means  $E(m,\eta)\neq E(m,-\eta)$. Now when $r\to \infty $, we observe that the energy levels  are almost constant. For $\Delta=0$ meV, Figures \ref{f3}(a,b,c) show  the presence of  zero energy levels in similar way to the results derived in our previous work \cite{Belouad20}. For the case of a gap $\Delta=100$ meV, Figures \ref{f3}(d,e,f) show that the energy levels possess an energy gap  $2\Delta$ between the valance and conduction bands. Additionally, 
  we notice that the energy levels verify the antisymmetry $ E(m,\eta)\neq E(-m, \pm\eta)$ and asymmetry $E (m,\eta)=-E (m,-\eta)$ relations.

\begin{figure}[!hbt]\centering
	\center
	\includegraphics[width=5.6cm]{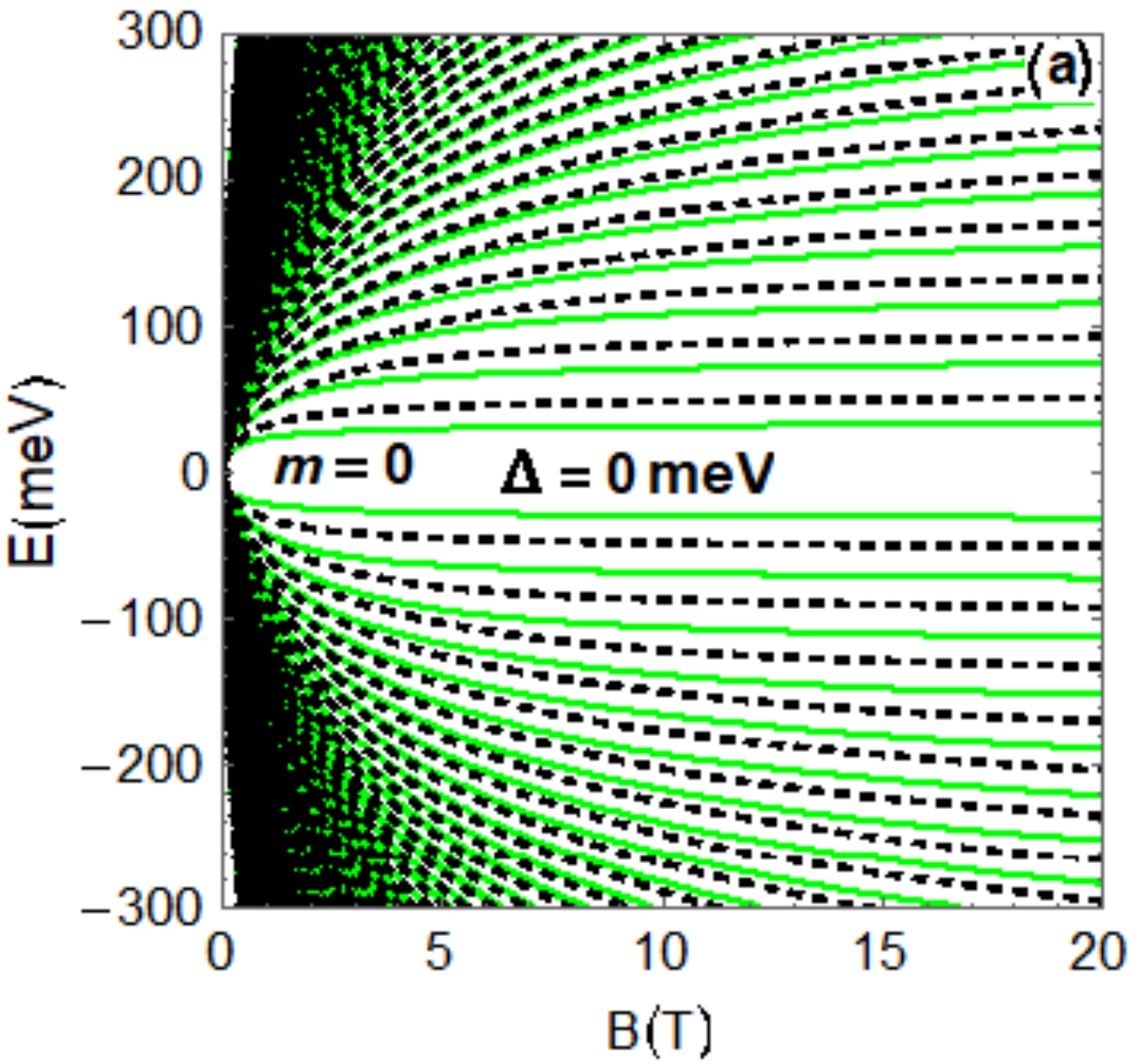}\includegraphics[width=5.6cm]{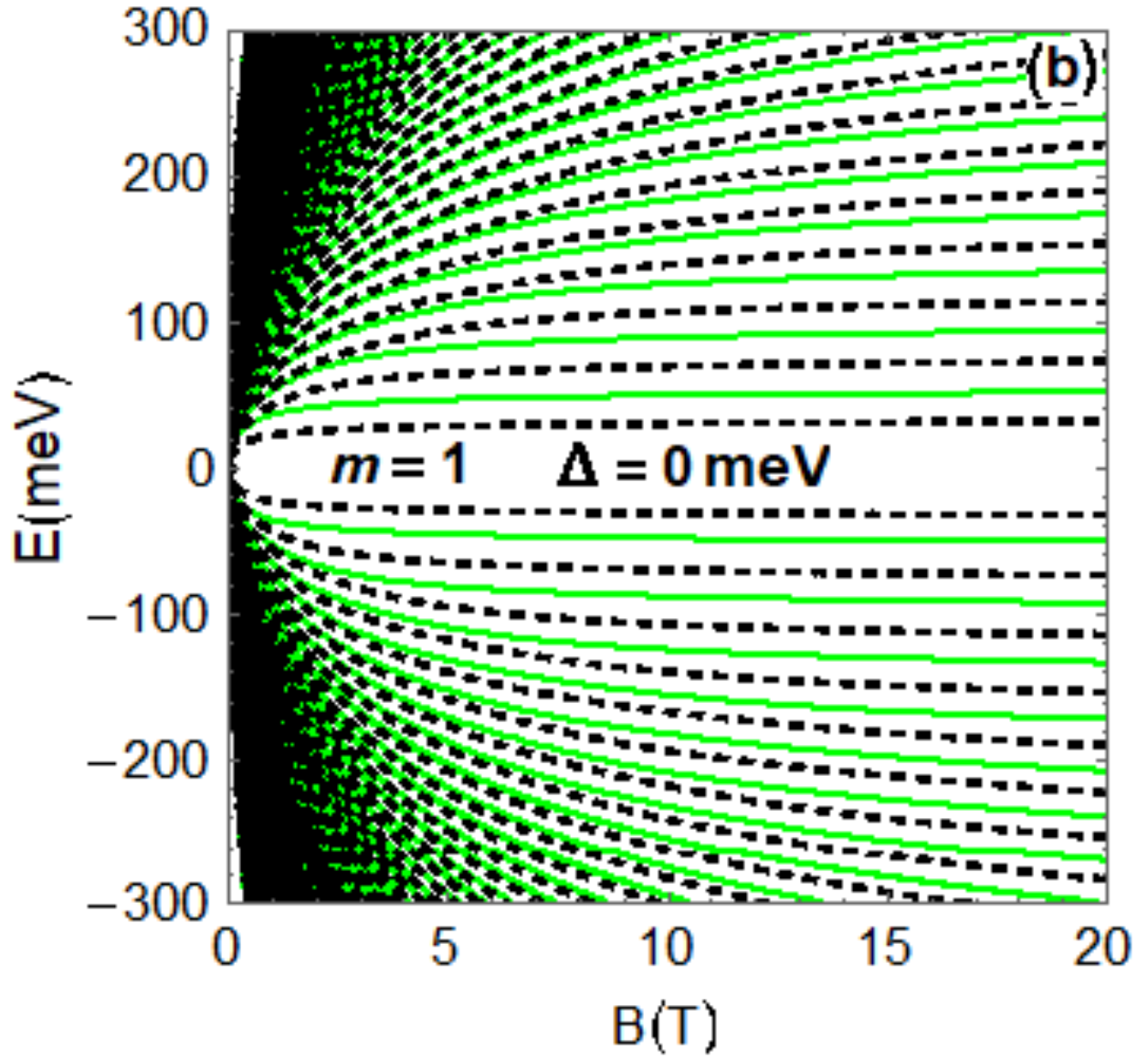}\includegraphics[width=5.6cm]{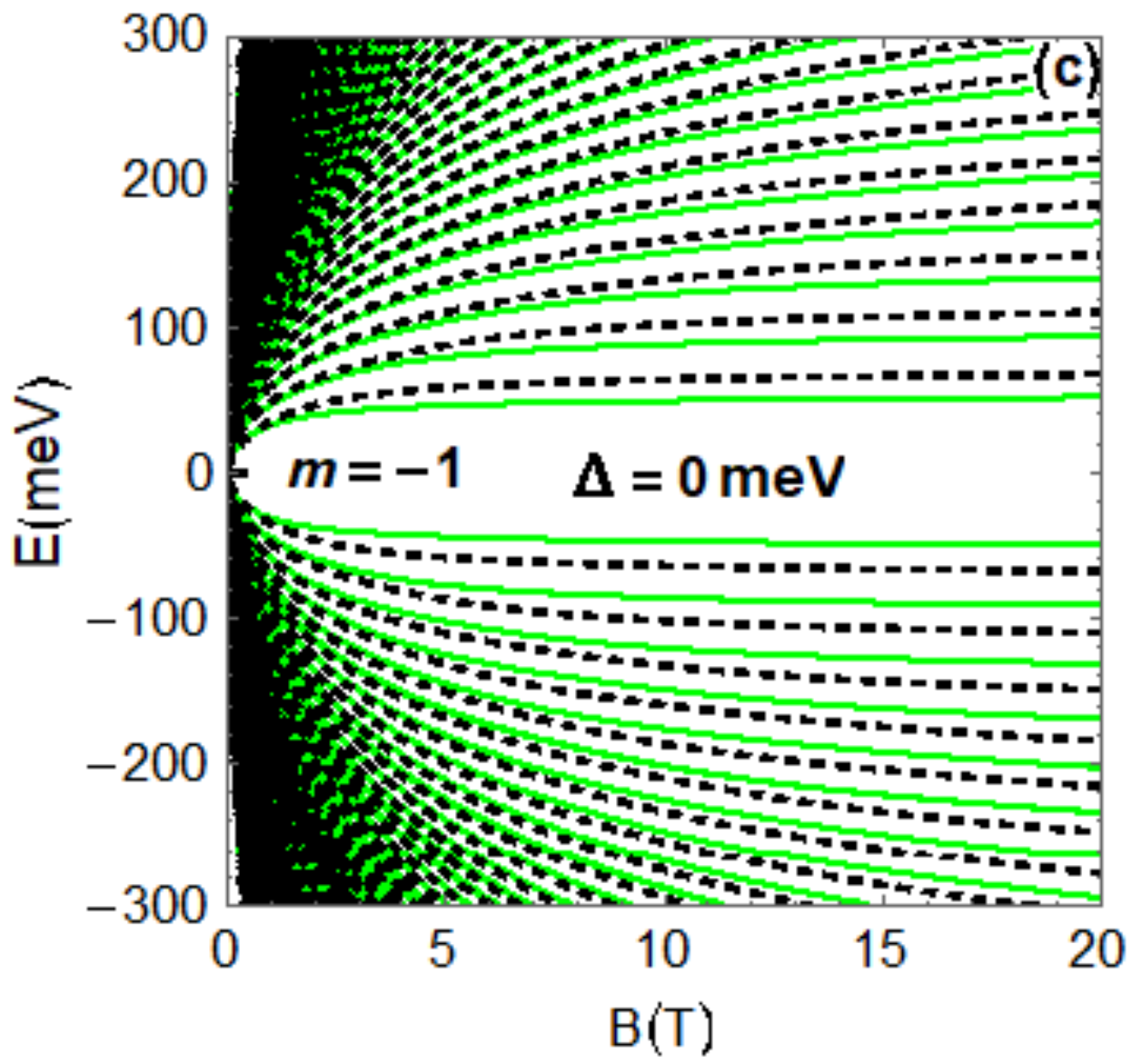}
	\includegraphics[width=5.6cm]{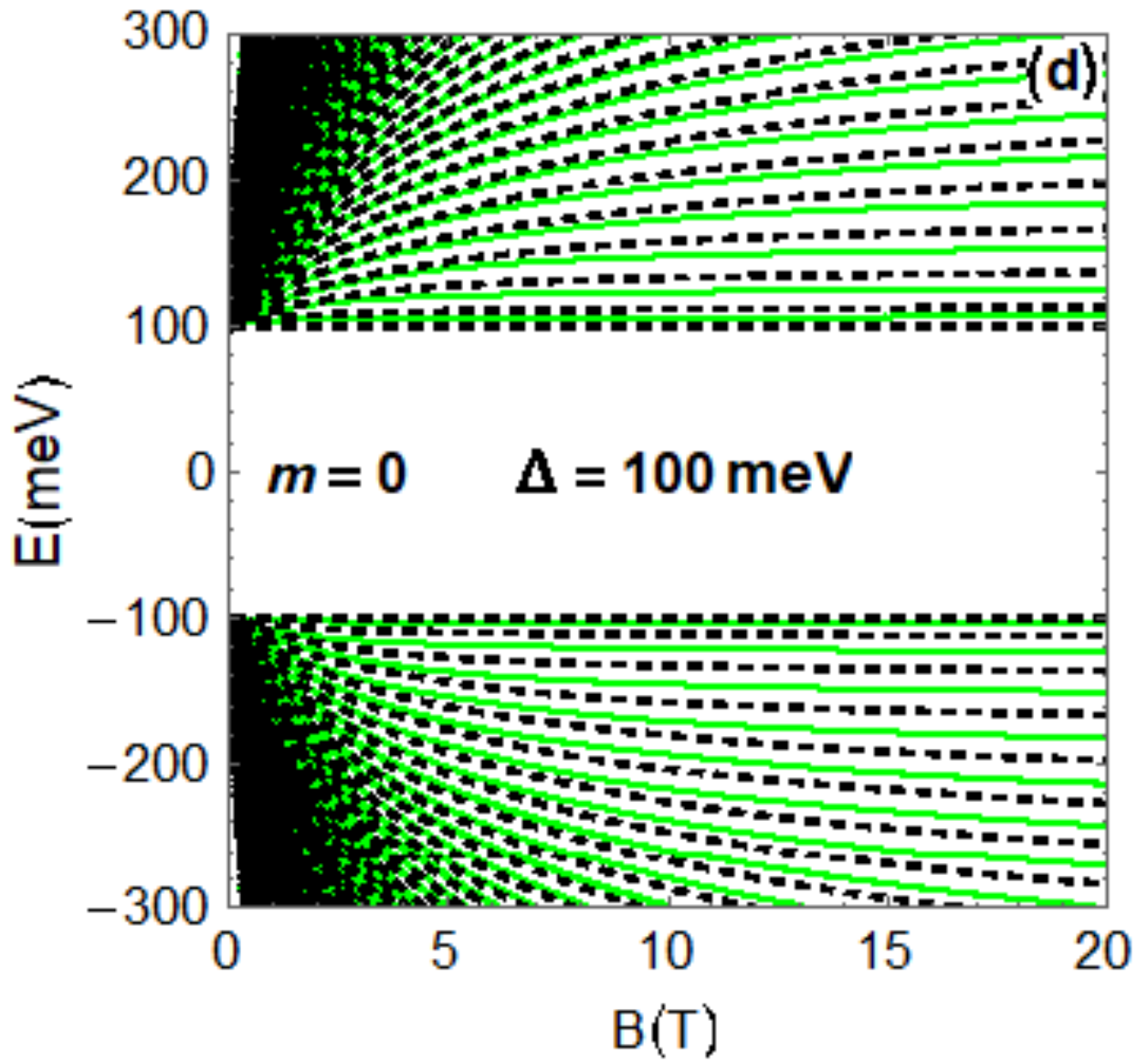}\includegraphics[width=5.6cm]{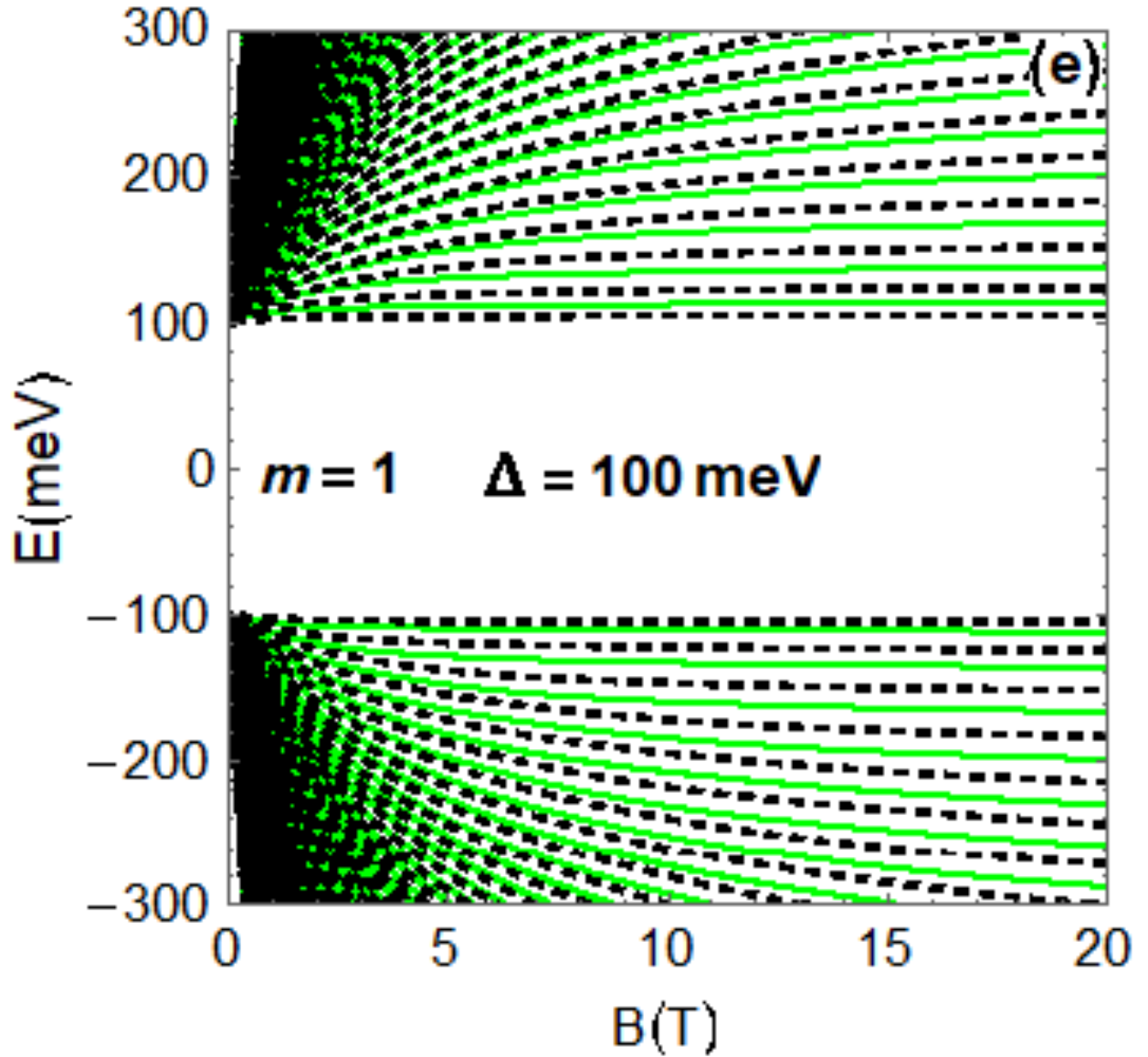}\includegraphics[width=5.6cm]{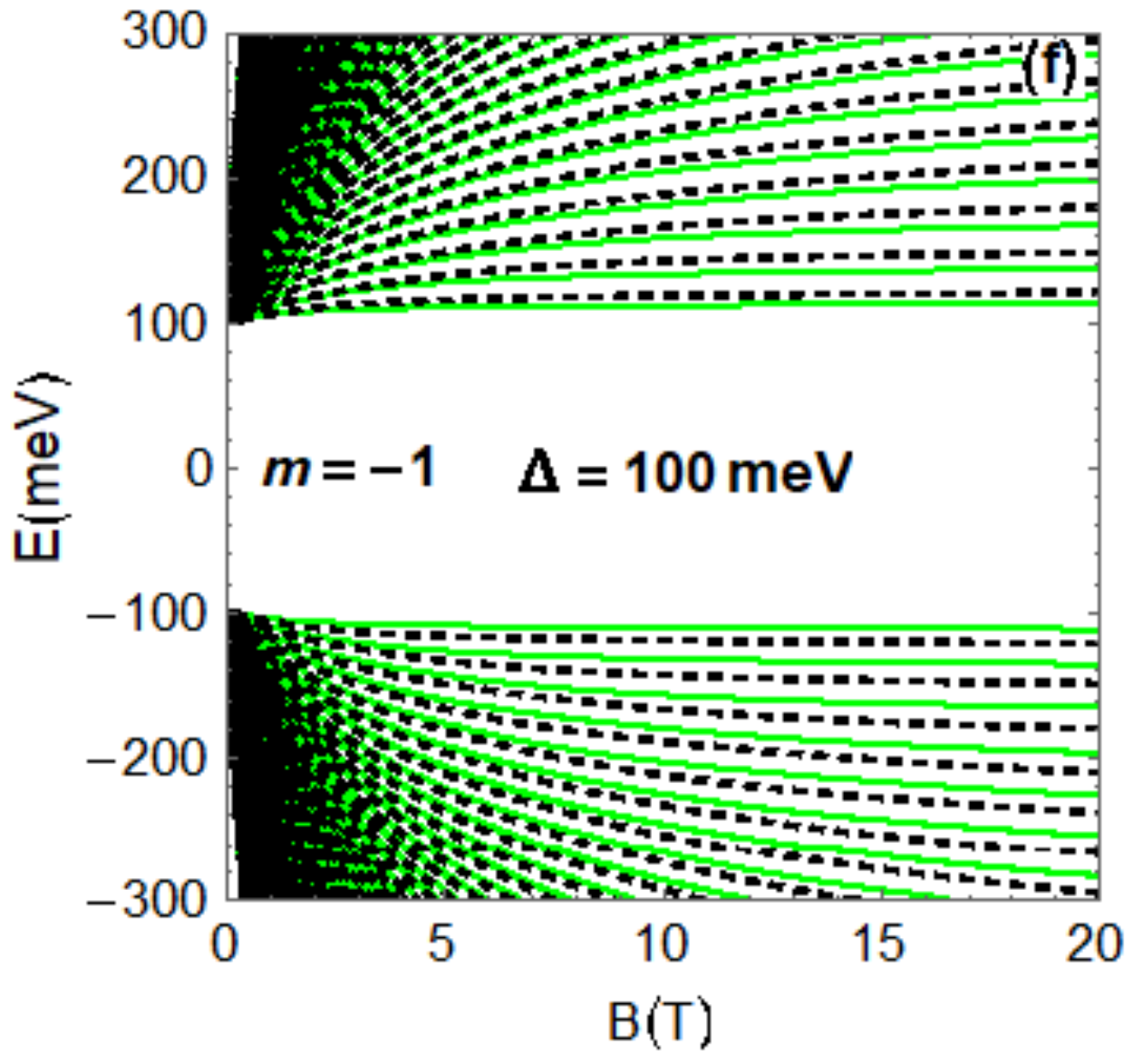}
	\caption{\sf (color online) Energy levels as a function of the magnetic field $B$ for $r_0=44.3$ nm, three values of angular momentum and two values of energy gap. (a): $m=0$, (b): $m=1$, (c): $m=-1$ for $\Delta=0$ meV. (d): $m=0$, (e): $m=1$, (f):   $m=-1$ for $\Delta=100$ meV, with green curves for $\eta=1$ and black dashed curves for $\eta=-1$. \label{f4}}
\end{figure}

In Figure \ref{f4}, we present 
the energy levels as a function of the magnetic field $B$   for $r_0=44.3$ nm, three values of angular momentum  $m$ and energy gap $\Delta$. More precisely, we have (a): $m=0$, $(b)$: $m=1$, (c): $m=-1$ for $\Delta=0$ meV and (d): $m=0$, (e): $m=1$,  (f): $m=-1$ for $\Delta=100$ meV. It is clearly seen that for a weak magnetic field ($B\to 0 $), there are many degenerate energy states corresponding to all angular momentum, for the valleys $K$ and $K'$, in the form of a continuous energy band  $E(B,\eta)= E(B,-\eta)
$\cite{Mirzakhani16,Belouad20}. On the other hand,
it is interesting to note that there are creation of gaps when $B$ increases as shown in Figures \ref{f4}(a,b,c) even in the absence of energy gap ($\Delta=0$ meV). Now we observe in  Figures \ref{f4}(d,e,f) that 
 the values $ \Delta=100$ meV increases the energy gap between  valence  and conduction bands. Then the energy gap can be used as a  tunable parameter to control and adjust the energy levels.

\begin{figure}[h!]
	\center
	\includegraphics[width=8.5cm]{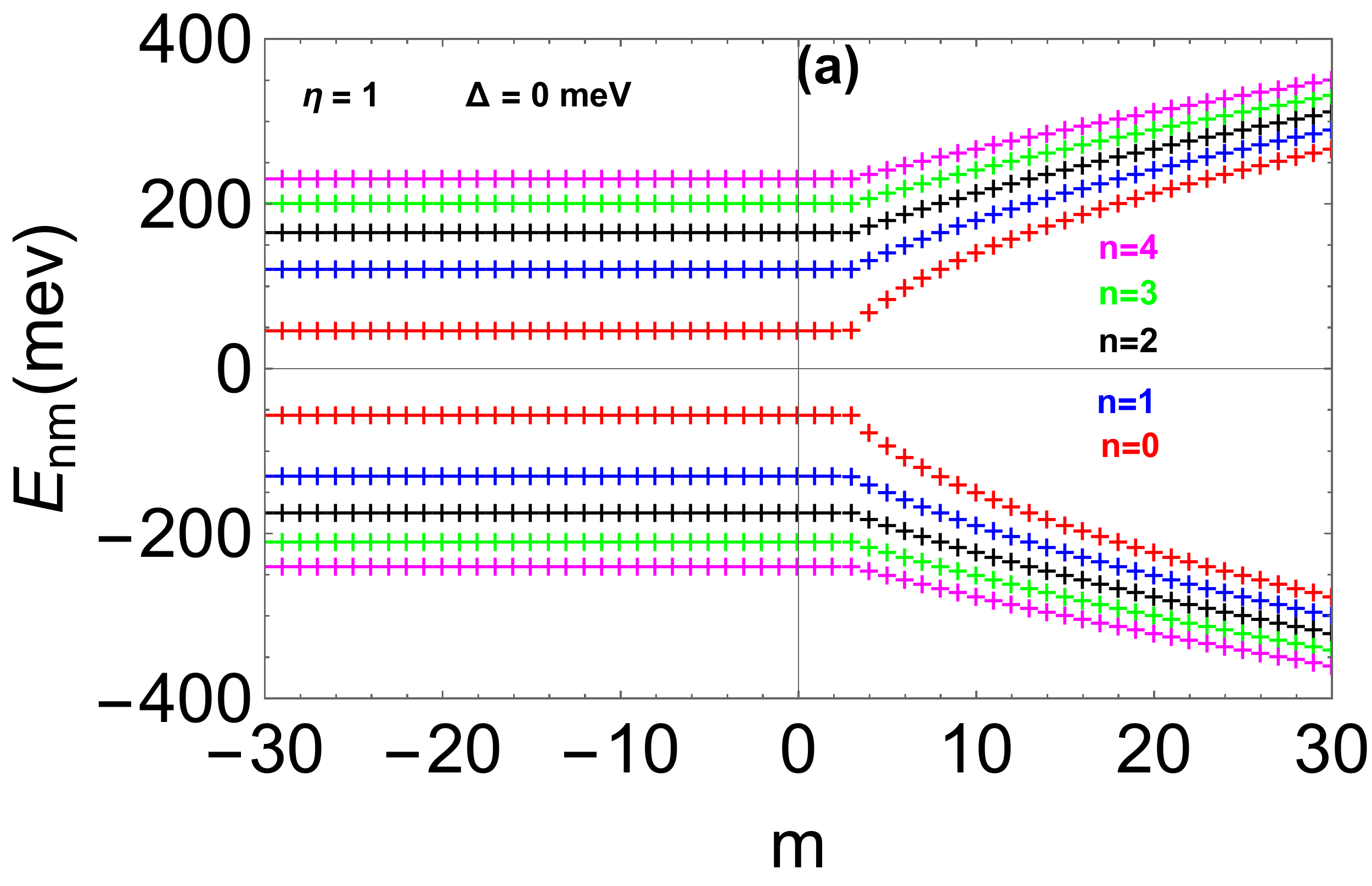}\includegraphics[width=8.5cm]{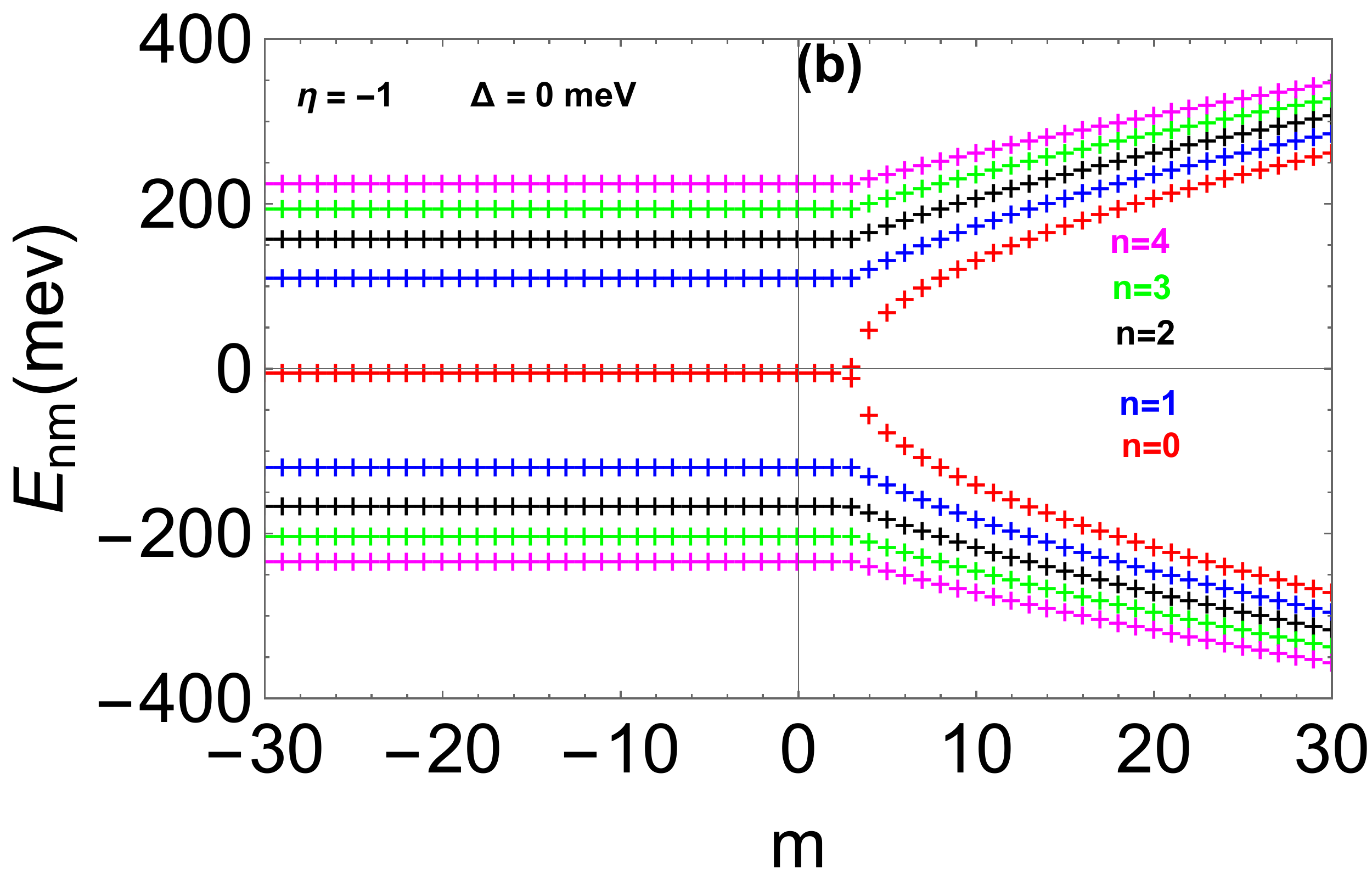}
	\includegraphics[width=8.5cm]{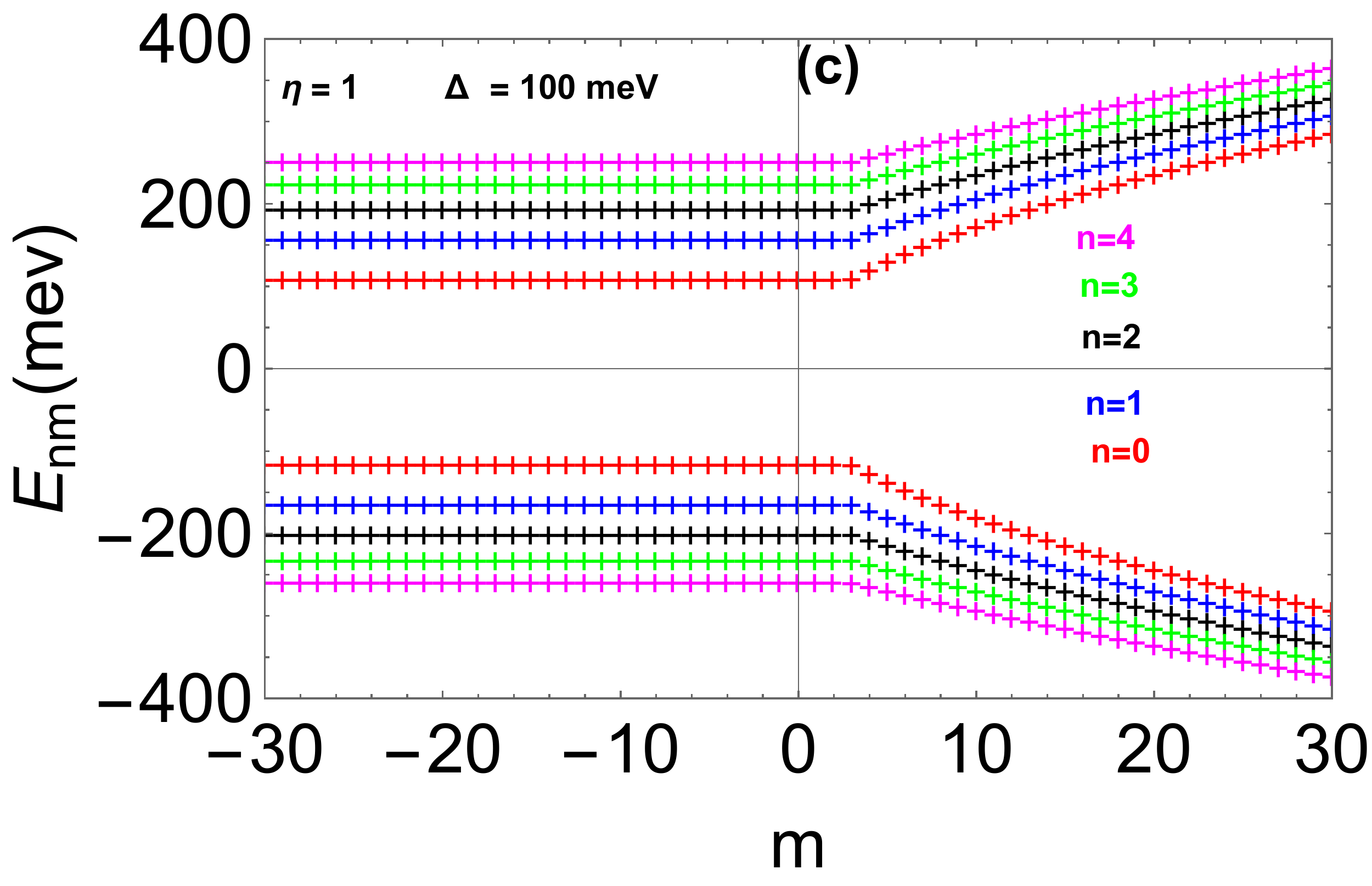}\includegraphics[width=8.5cm]{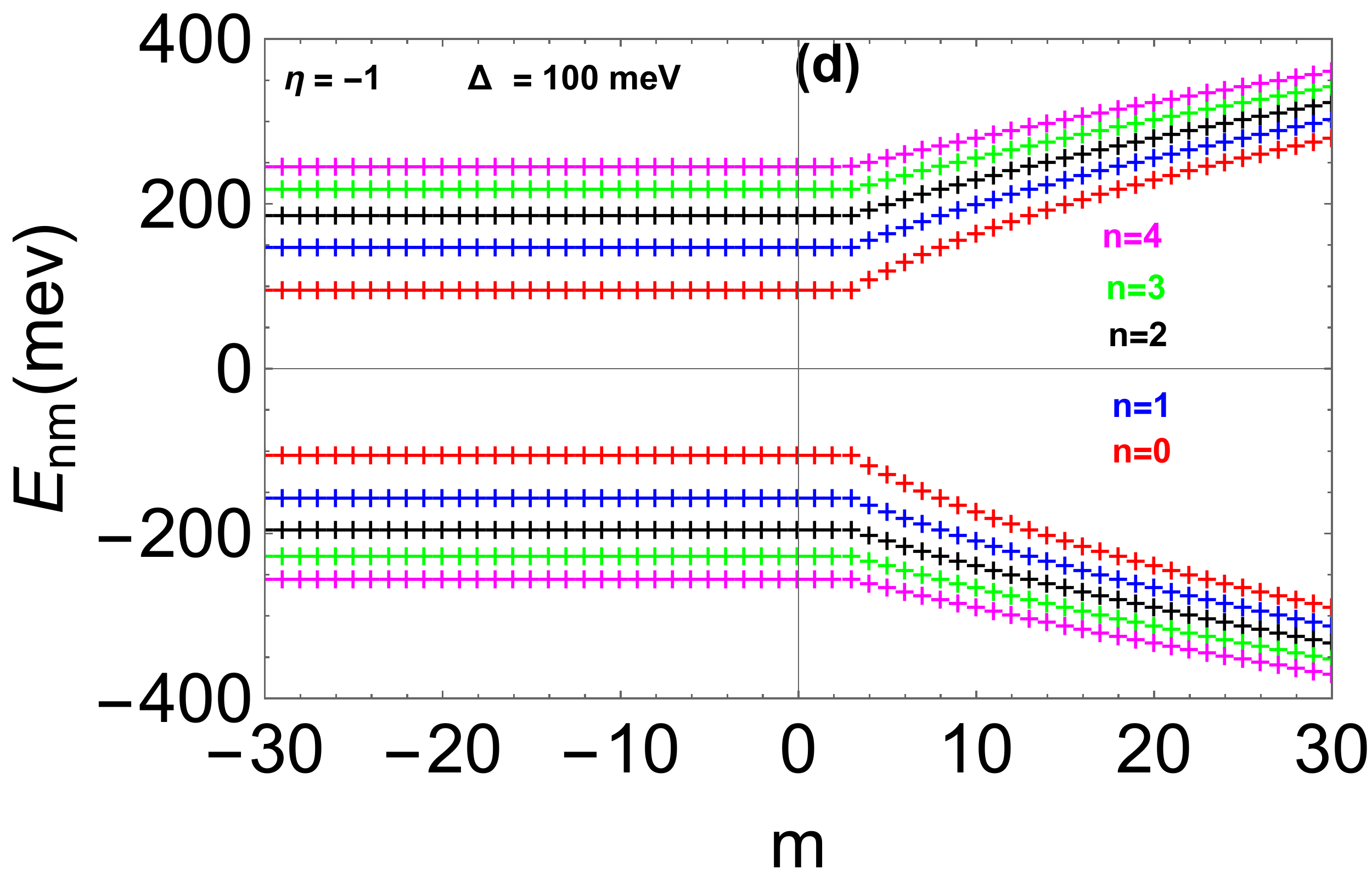}
	\caption{\sf (color online) Energy levels $E_{nm}$  as a function of the angular momentum $m$ for   $B=12$ T and five Landau levels $n=0, \cdots 4$. (a): $\Delta=0$ meV, (c): $\Delta=100$ meV for $\eta=1$ and (b): $\Delta=0$ meV, (d): $\Delta=100$ meV for $\eta=-1$.\label{f8}}
\end{figure}

Now let us see what happen for high magnetic field case. Indeed, 
by increasing $B$ we show that the degenerate levels for each $m$ are lifted due to the broken  symmetry, i.e.  $E(B,\eta)\neq E(B,-\eta)$. Consequently, we can find  an explicit expression for the energy levels involving Landau levels $n$ in addition to angular momentum $m$
\begin{equation}\label{e28}
E_{nm}=\pm E_0\sqrt{n+ \delta^2+ \frac{|m_{\text{eff}}|+m_{\text{eff}}+1+\eta}{2}}, \qquad n=0, 1,2, \cdots
\end{equation}
which can be derived by using \eqref{e20} and requiring the condition $a=-n$. In this case, the confluent hypergeometric function will be replaced by 
the Laguerre one \cite{Recher09}.
In Figure \ref{f8} we present the energy levels  $E_{nm}$  for $\Delta=0$ meV in  panels (a,b) and $\Delta=100$ meV in
 panels (c,d) with Landau levels $n=0, \cdots ,4$. We observe that $E_{nm}$ show an asymmetric behavior between the valence  and  conduction bands, i.e. $E(m,\eta)=-E (m,\eta)$ \cite{Belouad20}. In the absence of  energy gap ($\Delta=0$ meV), we notice that the energy level $n = 0$ is not a degenerate state for the  valleys
$K$ and $K'$ because for $m<4$ it is no-null for $\eta=1$ but null for $\eta=-1$.  While the other Landau levels  are degenerate states for these two valleys as shown in  panels (a,b). 
In the presence of  energy gap ($\Delta\neq0$),
it is clear from  panels (c,d) that  all the energy levels are degenerated for the two valleys and there is an increase in the energy gap between the  valence and conduction bands.

\begin{figure}[h]\centering
  \center
  \includegraphics[width=5.7cm]{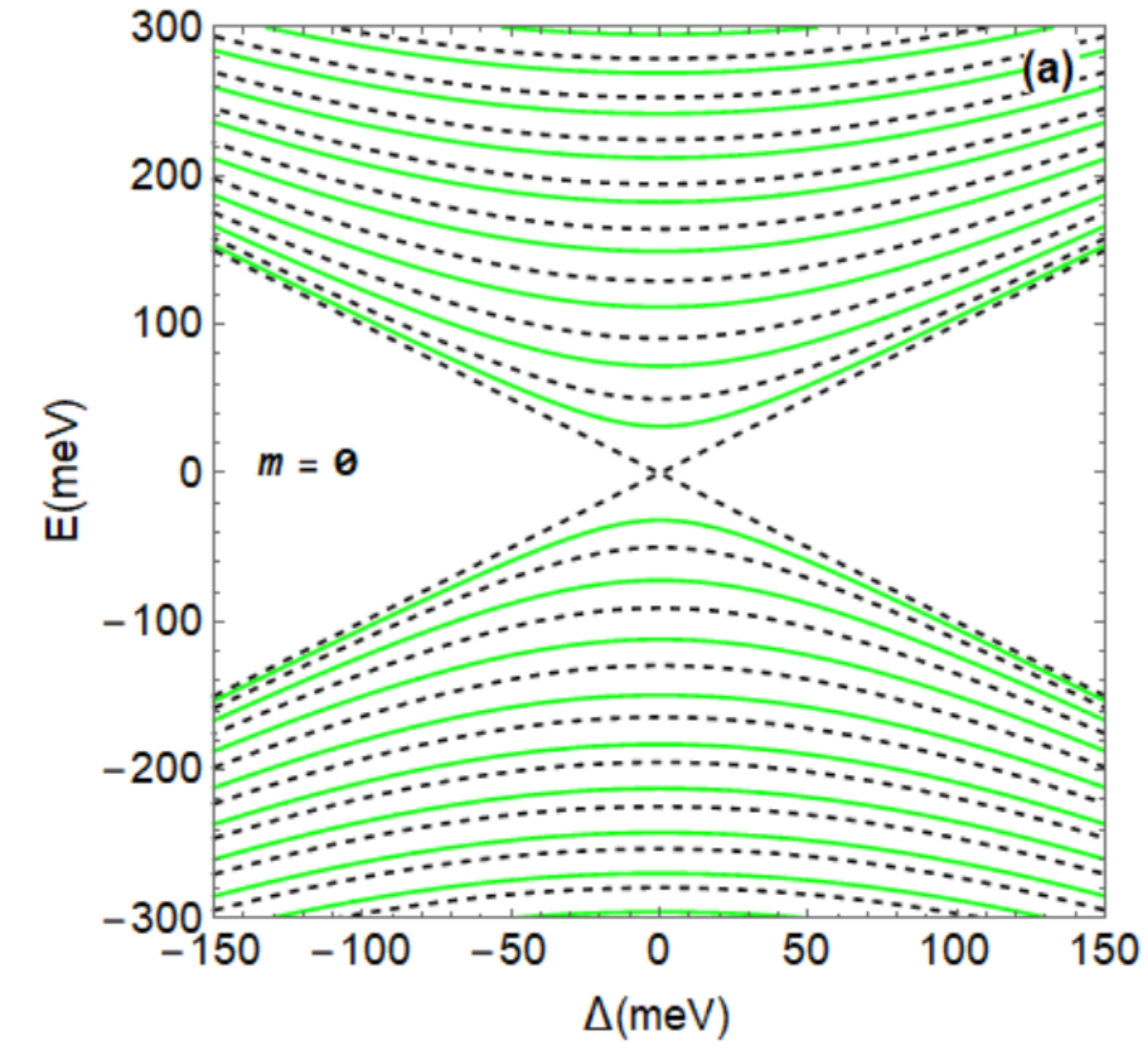}\includegraphics[width=5.7cm]{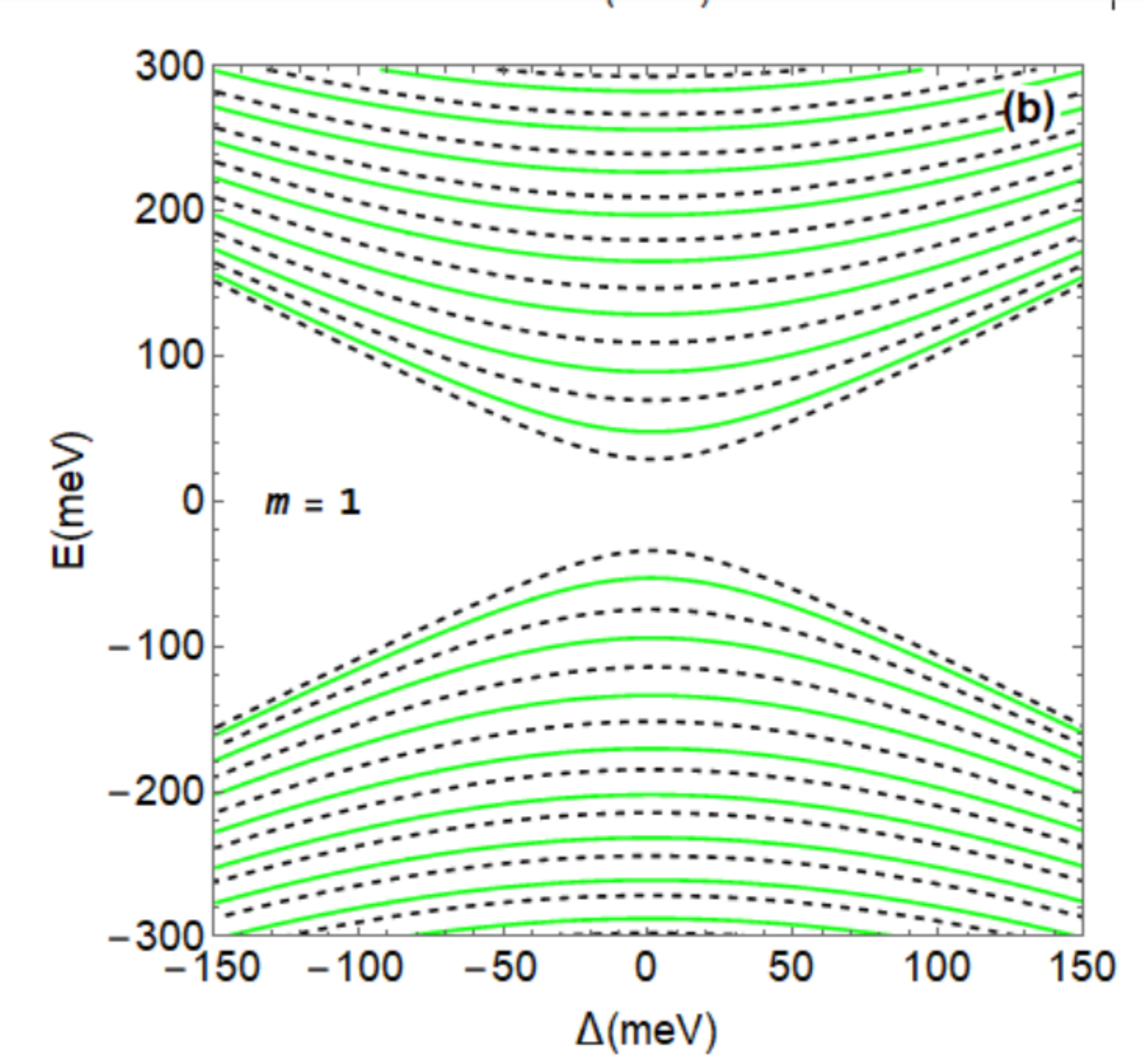}\includegraphics[width=5.7cm]{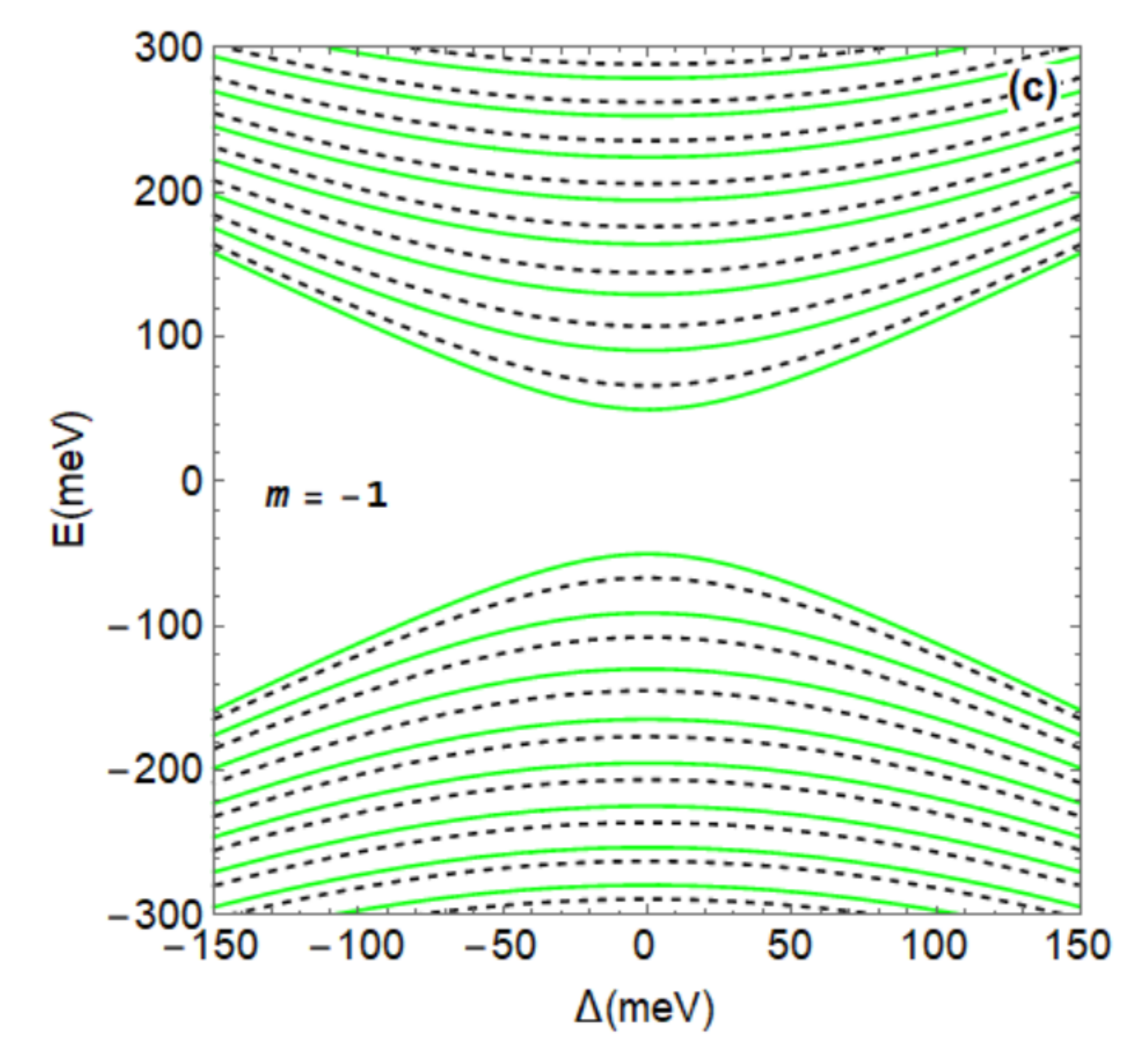}
  \caption{\sf (color online) Energy levels as a function of the energy gap $\Delta$ for $B=15.7$ T, $r_0=44.3$ nm and three values of angular momentum $m$.  (a): $m=0$, (b): $m=1$, (c): $m=-1$, with  green curves for $\eta=1$ and black dashed curves for $\eta=-1$. \label{f5}}
\end{figure}

The dependence of the energy levels on the energy gap $\Delta$ is shown in Figure \ref{f5} for a magnetic field  $B=15.7$ T, radius $r_0=44.3$ nm  and three values of angular momentum $m$ with (a): $m=0$, (b): $m=1$ and (c): $m=-1$. Note that the solid  and dashed lines correspond to the two valleys $K\ (\eta=1) $ and $K'\ (\eta=-1)$,  respectively.
We observe that the energy levels show a parabolic behavior with a minimum corresponds to $\Delta=0$ meV and satisfy two symmetry relations such that $E(\Delta, \eta) = -E(\Delta,\eta)$ and $E(\Delta, \eta,m=1)=E(\Delta,\eta,m=-1)$.
For $m=0$ in Figure \ref{f5}(a), it is clearly seen  the existence of the  state $\eta=-1$ (black dashed curves) inside the gap. On the other hand, for
$m=\pm 1$ in Figures \ref{f5}(b,c), we notice that the energy levels have a energy gap  even for $\Delta = 0$. This result is quantitatively similar to that found for a quantum ring consisting of a single layer of graphene \cite{Zarenia10}.

\begin{figure}[H]\centering
  \center
  \includegraphics[width=5.7cm]{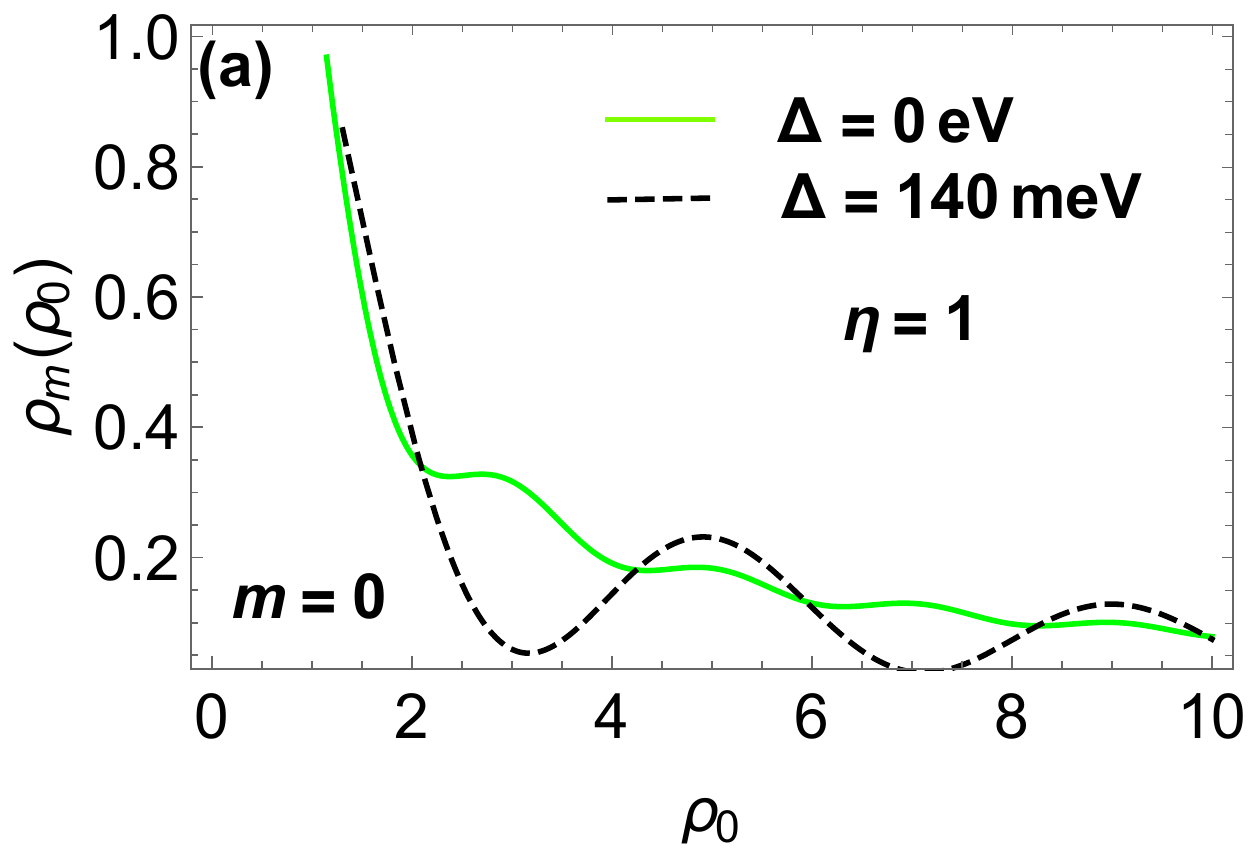}\includegraphics[width=5.7cm]{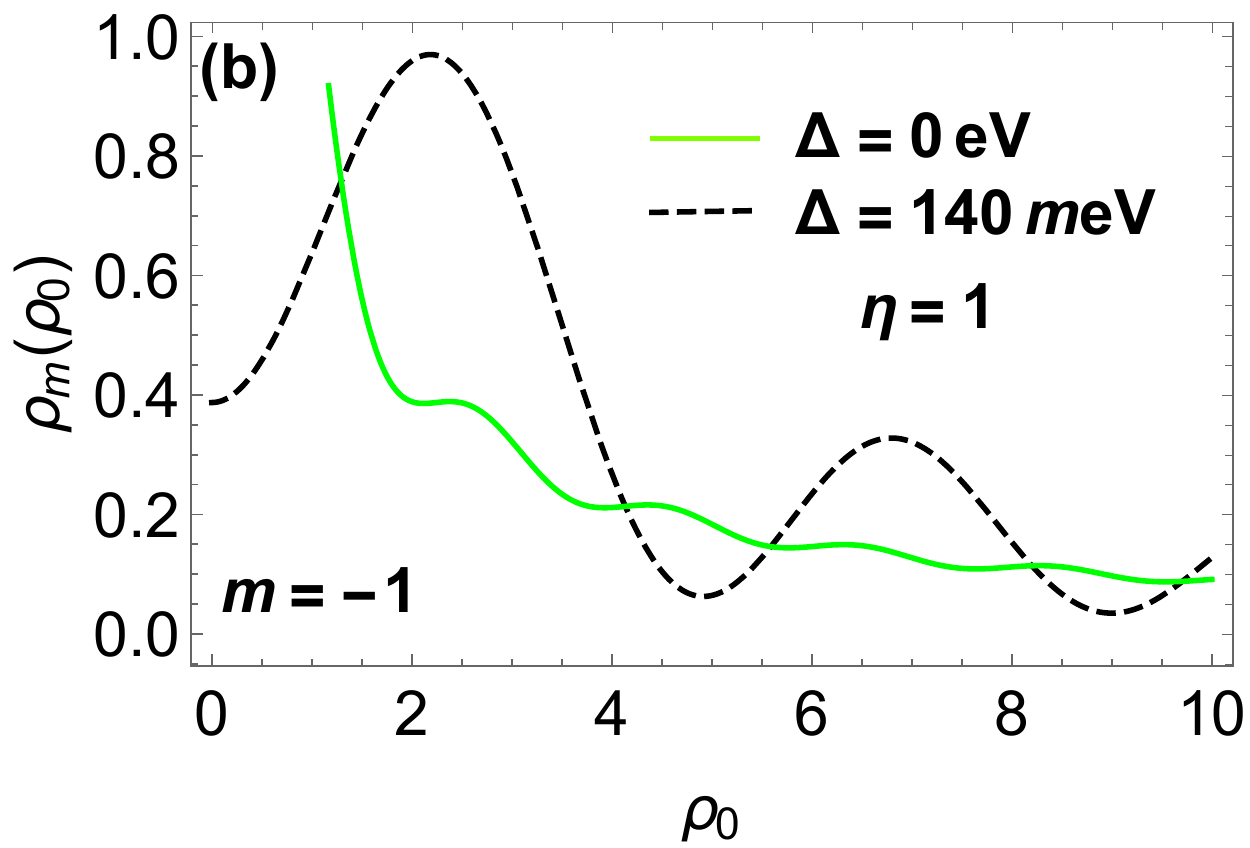}\includegraphics[width=5.7cm]{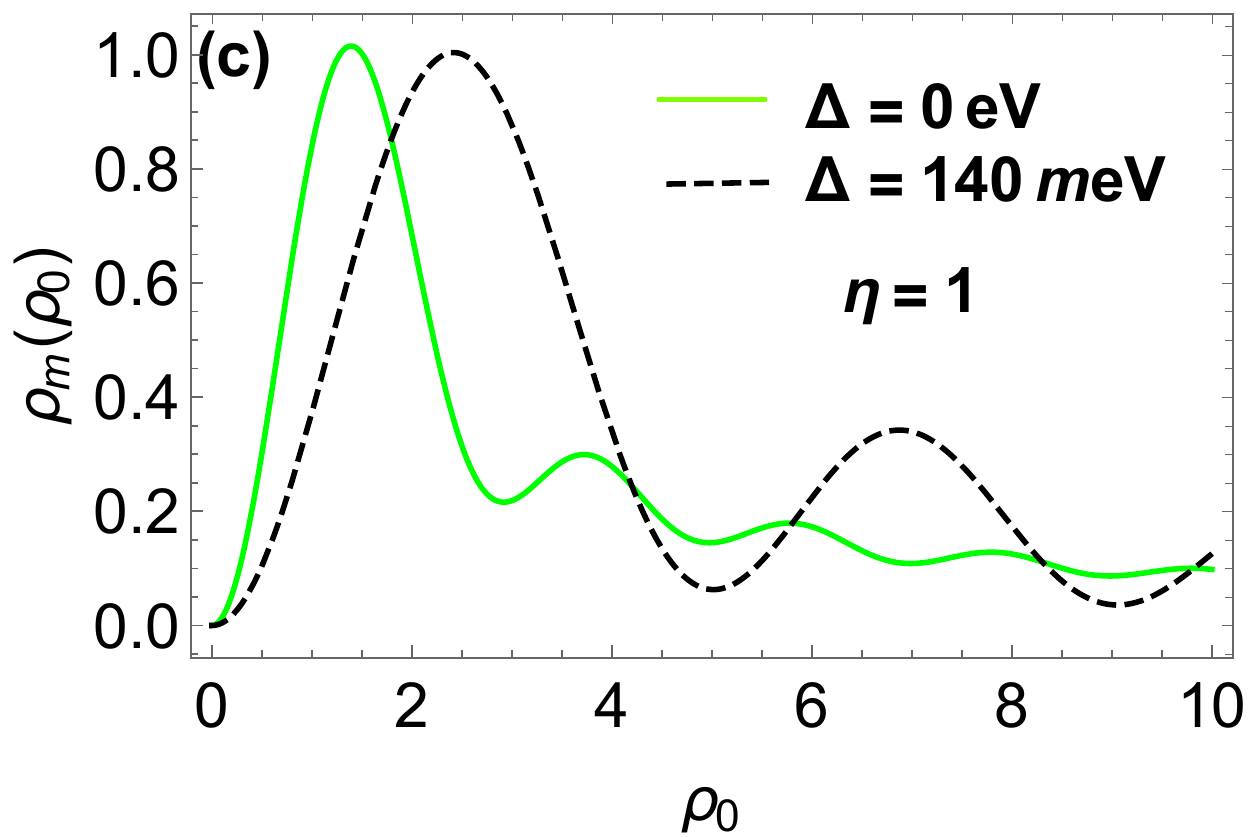}\\
  \includegraphics[width=5.7cm]{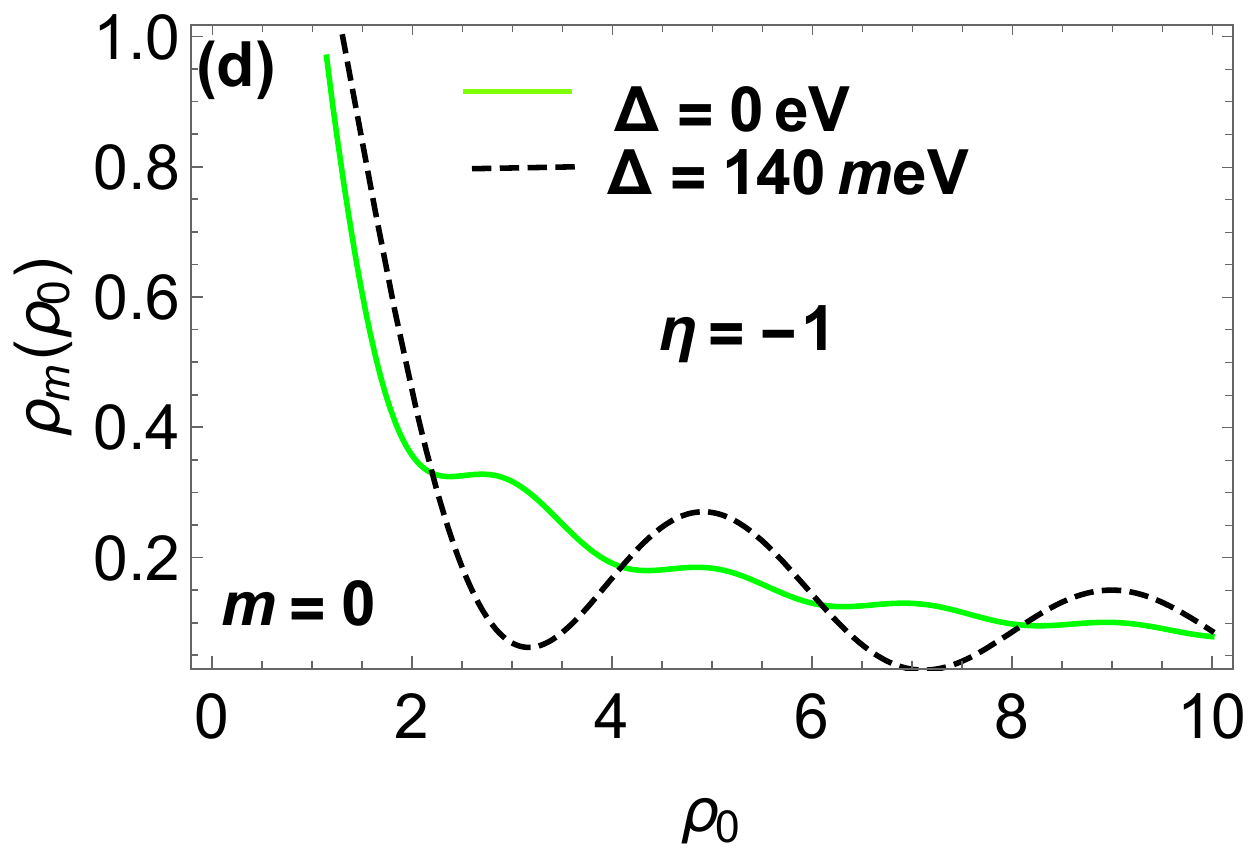}\includegraphics[width=5.7cm]{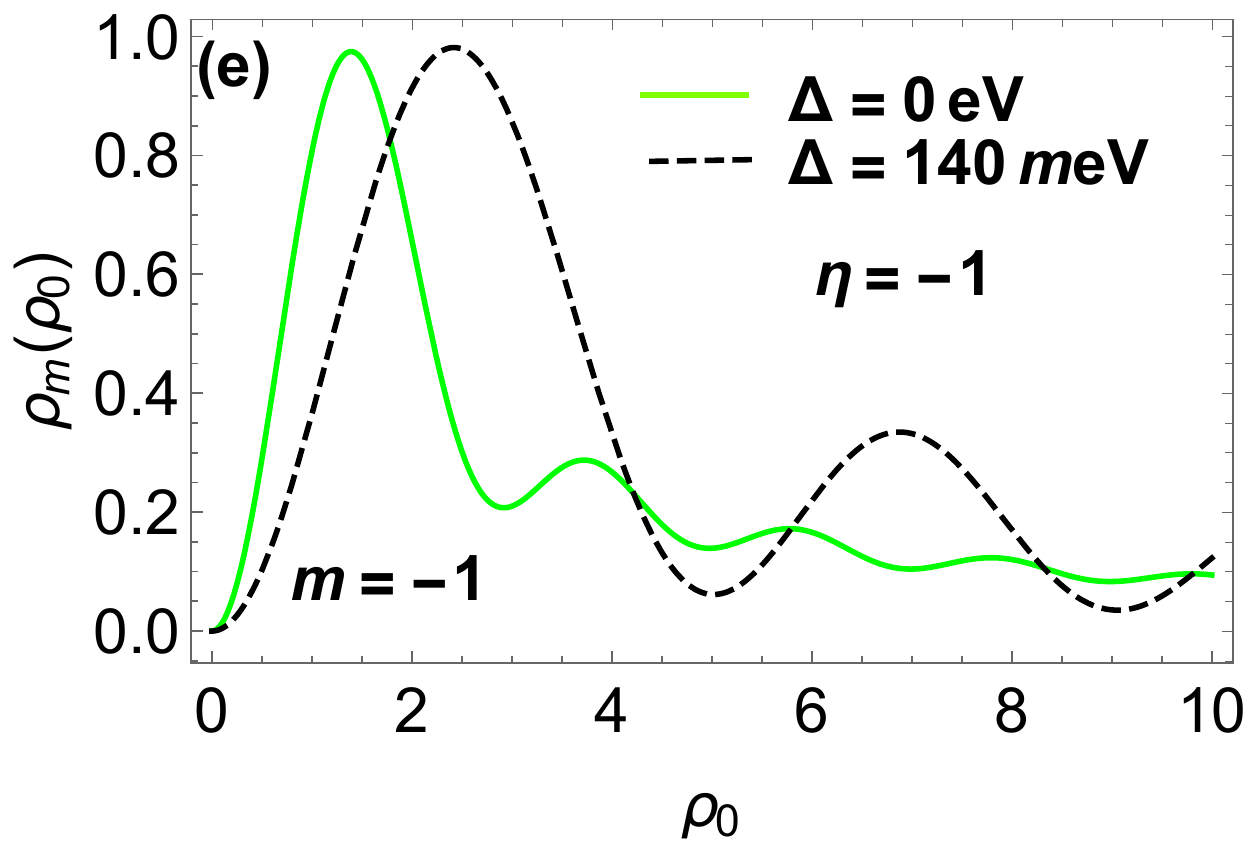}\includegraphics[width=5.7cm]{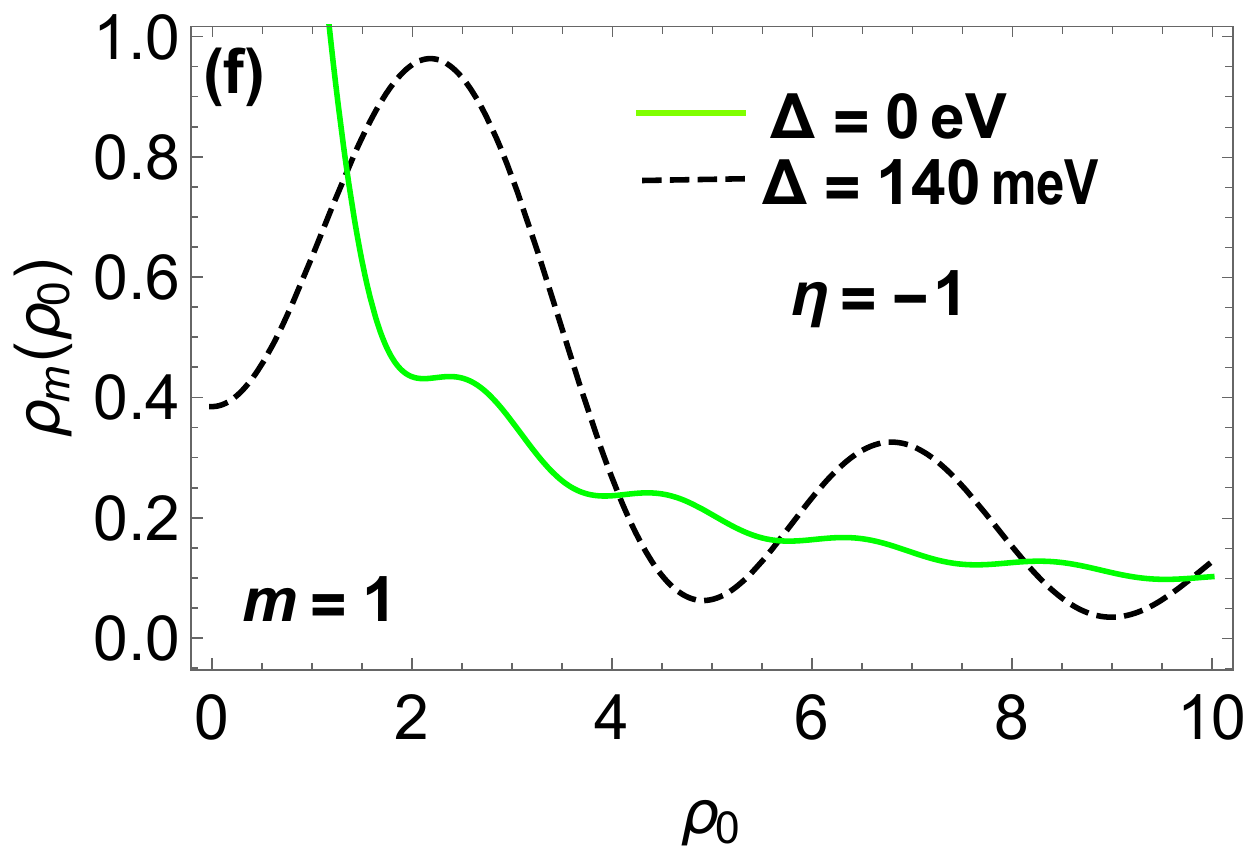}
  \caption{\sf (color online) Radial probability $\rho_m(\rho_0)$ as a function of the normalized radius  $\rho_0 =\frac{r_0}{l_B}$ of quantum dot  for $B=15$ T, $E=160$ meV, $m=0 ,-1,1$. (a,b,c): $\eta=1$, (d,e,f):  $\eta=-1$.  $\Delta=0$ meV: green curves and  $\Delta=140$ meV: black dashed curves.   \label{f7}}
\end{figure}

Figure \ref{f7} presents the radial probability $\rho_m(\rho_0)$ as a function of the normalized radius $\rho_0=\frac{r_0}{l_B}$ of quantum dot for $B=15$ T, $E=160$ meV and $m=0 ,-1,1$ with $\Delta=0$ meV (green) and  $\Delta=140$ meV (black dashed). We notice the symmetries $\rho_m(\rho_0,\eta)=\rho_m(\rho_0,-\eta)$ for $m=0$ as shown in panels (a,d) and $\rho_m(\rho_0,\eta)=\rho_{-m}(\rho_0,-\eta)$ for $m\neq0$  as presented in panels (b,f) and (c,e).  We observe that  when $\rho_0$ decreases $\rho_m(\rho_0)$ tends to a maximum value near $\rho_0 =2$.  On the other hand, we have zero radial
probability in the vicinity of $\rho_0=0$ for $m=1$ where $\eta=1$ see panel (c) and for $m=-1$ where $\eta=-1$ see panel (e). In addition, the radial probability approximately oscillates with damping as $\rho_0$ increases \cite{Belouad20}. We notice that the presence of energy gap ($\Delta\neq0$) causes a shift of $\rho_m(\rho_0)$  when one moves away from the center of quantum dot and also causes a decrease in the damping of  oscillations observed in the case where $\Delta=0$.

\begin{figure}[!hbt]\centering
	\center
	\includegraphics[width=17cm]{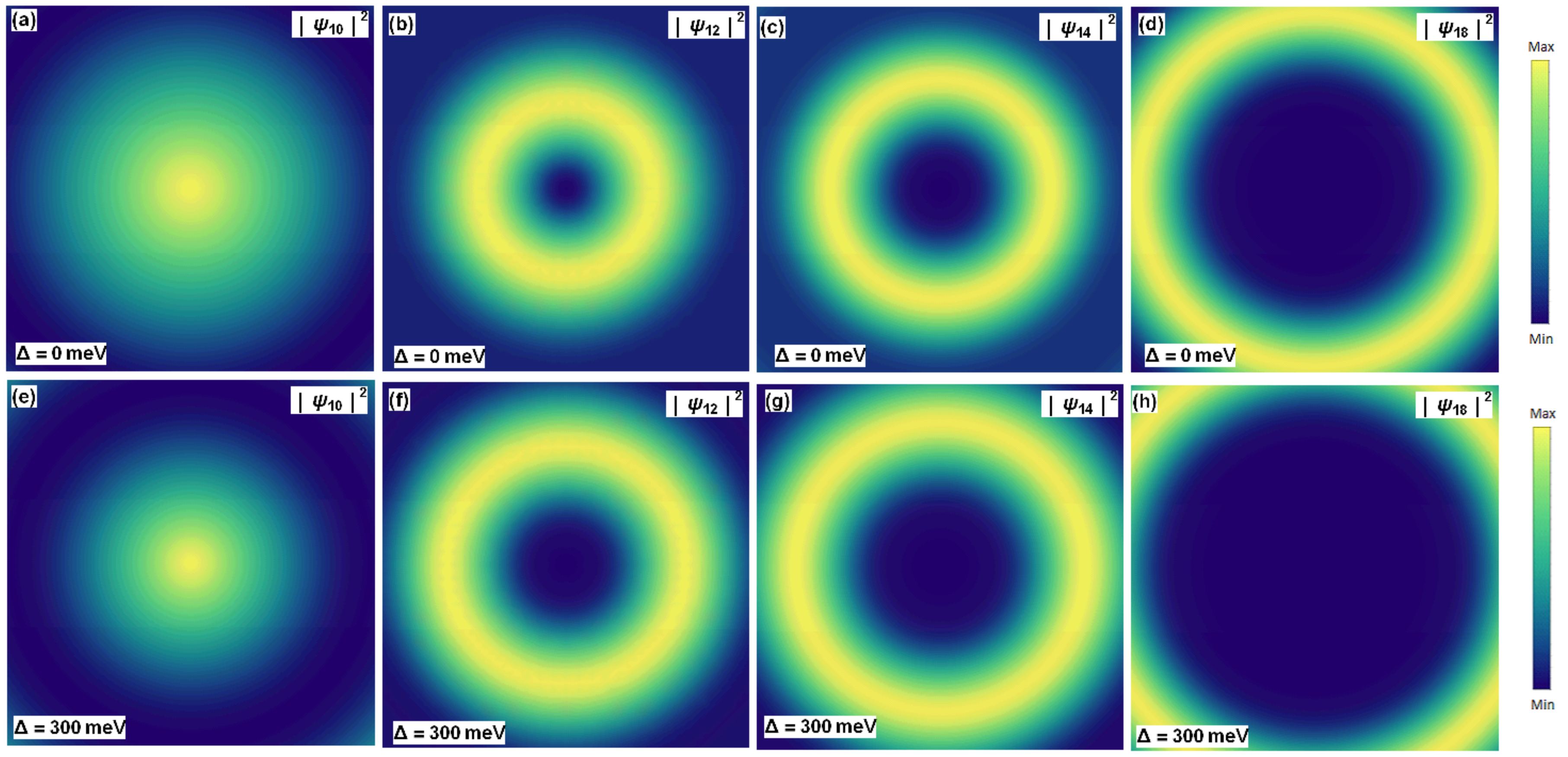}
	
	\caption{(color online) \sf Spatial density $|\psi_{nm}|^{2}$ in the vicinity of the quantum dot for  $\eta=\pm1$, $B=15.7$ T. (a): ($n=1$, $m=0$), (b): ($n=1$, $m=2$), (c): ($n=1$, $m=4$), (d): ($n=1$, $m=8$) for $\Delta=0$ meV and (e): ($n=1$, $m=0$), (f): ($n=1$, $m=2$), (g): ($n=1$, $m=4$), (h): ($n=1$, $m=8$) for $\Delta=300$ meV. \label{f6}}
\end{figure}

Figure \ref{f6} shows the electron density of charge carriers in 
the quantum dot for $B=15.7$ T
and some particular values of the quantum numbers $n$ and $m$.
We observe that
the electron density has  maxima at the center of quantum dot for $m=0$ \cite{ Schulz15}. Such maxima decrease when the  energy gap ($\Delta\neq 0$) is considered as it can be seen clearly by comparing panels (a,b,c,d) for
$\Delta=0$ meV and panels (e,f,g,h) for $\Delta=300$ meV. On the other hand, for
the case  $m\neq 0$, the electron density shows a minima at the center of quantum dot  \cite{ Schulz15}, which  increase  for $\Delta\neq 0$ as can be seen by looking at  panels (b,c,d) for $\Delta=0$ meV and panels (f,g,h) for $\Delta=300$ meV.

\section{Conclusion}

We have studied the electronic properties of a system consisting of a quantum dot surrounded by a graphene sheet with energy gap $\Delta$ in the presence of a magnetic field. By solving the  Dirac equation with two bands, in the vicinity of the two valleys $K$ and $K'$, we have derived the eigenspinors. Thanks to the boundary condition, we have obtained an analytical formula including all physical parameters characterizing our system to describe the energy levels.

We have shown that the energy levels present an asymmetry between the valence
and the conduction bands, i.e.
$E(B,\Delta,m,\eta)=-E(B,\Delta,m,\eta)$.
For a very small size of  QD ($r_0\to 0$) the energy levels correspond to $K$ ($\eta=1$) and $K'$ $\eta=-1$) valleys are degenerate, that is to say
$E(m,\eta)=E(m,-\eta)$. It was observed that  when the size becomes large the degenerate of the valleys $K$ and $K'$ is broken
$E(m,\eta)\neq E(m,-\eta)$.
We have shown that for a weak magnetic field $B\to 0$, there are many states of degenerate energy corresponding to all the angular moments $m$ for the valleys $K$ and $K'$ in the form of a band of continuous energy. By increasing the magnetic field, the degenerate levels for each $m$ are raised because the symmetry is broken. In each representation of the energy levels shows that the introduction of energy gap $\Delta\neq0$ leads to an increase in the gap  between the valence and conduction bands.  

As far as  the radial probability $\rho_m (\rho_0)$ is concerned,  we have obtained two symmetries such that $\rho_m(\rho_0,\eta)=\rho_m(\rho_0,-\eta)$ for zero angular momentum $m=0$ and  $\rho_m(\rho_0,\eta)=\rho_{-m}(\rho_0,-\eta)$ for a non-zero angular momentum $m \neq0$. 
We have shown that
the radial probability presents a maximum when the normalized radius $\rho_0$ 
of quantum dot tends towards the value 2. It was noticed that the introduction of  energy gap ($\Delta \neq0$)  decreases the effect of the damping provided on the oscillation of $\rho_m (\rho_0)$ in comparison with the case where $\Delta =0$. Furthermore, we have shown that the electron density presents  maxima at the center of quantum dot for $m=0$ whereas for $m\neq 0$ it has minima.

\section*{Acknowledgment}
The generous  support provided by the Saudi Center for Theoretical Physics (SCTP) is highly appreciated by all authors.

\end{document}